\documentclass[twocolumn,tighten,times]{aastex62}
\usepackage{txfonts}
\usepackage{newlfont}

\shorttitle{$H_0$ Tension and the Phantom Regime}
\shortauthors{El-Zant, El Hanafy  \& Elgammal}

\begin{document}

\title{\uppercase{$H_0$ Tension and the Phantom Regime:\\ A Case Study In Terms of an Infrared \lowercase{$f$}$(T)$ Gravity}}

\author{Amr El-Zant}
\affiliation{Centre for theoretical physics, the British University in Egypt, 11837 - P.O. Box 43, Egypt}

\author[0000-0002-0097-6412]{Waleed El Hanafy}
\affiliation{Centre for theoretical physics, the British University in Egypt, 11837 - P.O. Box 43, Egypt}
\affiliation{Egyptian Relativity Group (ERG), Cairo University, Giza 12613, Egypt}

\author{Sherif Elgammal}
\affiliation{Centre for theoretical physics, the British University in Egypt, 11837 - P.O. Box 43, Egypt}

\correspondingauthor{Waleed El Hanafy}
\email{waleed.elhanafy@bue.edu.eg}


\begin{abstract}
We propose an $f(T)$ teleparallel gravity theory including a torsional infrared (IR) correction. We show that the governing Friedmann's equations of a spatially flat universe include a phantom-like effective dark energy term sourced by the torsion IR correction. As has been suggested, this phantom phase does indeed act as to reconcile  the tension between local and global measurements of the current Hubble value $H_0$. The resulting cosmological model predicts an electron scattering optical depth $\tau_e\thickapprox 0.058$ at reionization redshift $z_{re} \sim 8.1$,  in agreement with observations. The predictions are however in contradiction with baryon acoustic oscillations (BAO) measurements, particularly the distance indicators. We argue that this is the case with any model with a phantom dark energy model that has effects significant enough at redshifts $z \lesssim 2$ as to be currently observable. The reason being that such a scenario introduces systematic differences in terms of distance estimates in relation to the standard model; e.g., if the angular diameter distance to the recombination era is to be kept constant while $H_0$ is increased in the context of a phantom scenario, the distances there are systematically overestimated to all objects at redshifts smaller than recombination. But no such discrepancies exist between $\Lambda$CDM predictions and current data for $z \lesssim 2$.

\end{abstract}



\section{Introduction}\label{Sec:1}
Cosmological observations clearly confirm that our universe has speeded up its expansion as of a few billion years ago~\cite{Riess:1998AJ, Perlmutter:1999ApJ}, with  transition redshift $0.67\lesssim z_{tr}\lesssim 0.87$~\cite{Farooq:2016zwm}. In the context of general relativity (GR), explaining this
phenomenon requires the introduction of a cosmological constant or a negative pressure component in the field equations, referred
to as the dark energy. Several recent analyses, c.f. ~\cite{Ade:2015xua}, show that this component represents $\sim 69 \%$ of the energy density in the universe. The complementary components consist of $\sim 26\%$ and pressureless dark matter $\sim 5 \%$ baryons.

Since the dark energy effects are felt  on the cosmic scales, they are naturally tied to the gravitational interaction and its description.
In GR this involves the  field equations
\begin{equation}\label{GR-gravity}
    \frac{1}{\kappa^2} \mathfrak{G}_{\mu\nu}= \mathfrak{T}_{\mu\nu},
\end{equation}
where $\mathfrak{G}_{\mu\nu}$ is Einstein tensor, $\mathfrak{T}_{\mu\nu}$ is the energy-momentum tensor of the matter components;  the coupling constant $\kappa$, in the natural units ($c=\hbar=k_{B}=1$), can be related to the Newtonian constant $G$ by $\kappa=\sqrt{8\pi G}$.
To explain the late cosmic acceleration, one should represent the dark energy component $\mathfrak{T}^{DE}_{\mu\nu}$ as an additional term in Einstein's field equations as
\begin{equation}\label{Modified-gravity}
    \frac{1}{\kappa^2_{eff}} \mathfrak{G}_{\mu\nu}= \mathfrak{T}_{\mu\nu}+\mathfrak{T}^{DE}_{\mu\nu}\equiv \mathfrak{T}^{eff}_{\mu\nu},
\end{equation}
where the effective coupling constant $\kappa_{eff}$ reduces to the constant $\kappa$ at the GR limit and $\mathfrak{T}^{eff}_{\mu\nu}$ is the total energy-momentum tensor. The additional term $\mathfrak{T}^{DE}_{\mu\nu}$ in Eq. (\ref{Modified-gravity}) can be sourced either by matter (physical) or gravitational (geometrical) sectors. As noted by \cite{Sahni:2006pa}, the field equations (\ref{Modified-gravity}) put both physical and geometrical dark energies on equal footing. However, they provide different physical descriptions in some scenarios. For example, non-singular bounces have been studied as alternatives to big-bang, whereas the null energy condition should be violated. These have been investigated using effective field theory techniques by introducing matter fields which violate the null energy condition \cite{Cai:2007qw, Cai:2007zv}. On the contrary, using the modified gravity the null energy condition is violated effectively (gravitational sector) keeping the matter sector consistent with the null energy condition, c.f. \cite{Cai:2011tc,Bamba:2012vg,ElHanafy:2017sih} (see also the review \cite{Nojiri:2006ri}). In short, since the gravitational sector does not represent a physical matter field, we can exchange a particular exotic matter fields by some modified gravity without worry about the energy conditions.\\

(\textbf{i}) In the former case, the cosmological constant $\Lambda$ is the simpler scenario for the dark energy. This constant is equivalent to  a negative pressure term in Friedmann equations with equation of state parameter fixed to a value $w_\Lambda=-1$, allowing the universe to perform a transition from a decelerated expansion epoch dominated by cold dark matter (CDM) to an accelerated expansion dominated by $\Lambda$-dark energy, in agreement with observations. Although this  $\Lambda$CDM model fits well with a wide range of observations, it lacks adequate theoretical underpinning. Indeed, it entails several puzzling issues; e.g. the cosmic coincidence problem and the enormous discrepancy between its theoretical and observational values \cite{Weinberg:1988cp,Carroll:2000fy}. On the other hand, an alternative to the cosmological constant  consists of dynamical dark energy,  akin to inflaton fields,  which can be described as a canonical scalar field $\phi$ minimally coupled to gravity with fixed or dynamical equation of state parameter $-1 <w_\phi < -1/3$.

(\textbf{ii})  From geometrical point of view, there are three objects that can be used to describe deviations from Minkowski spacetime due to presence of a gravitational field, curvature $R$, torsion $T$ and non-metricity $Q$. When $\mathfrak{T}^{DE}_{\mu\nu}$ is sourced by the gravitational sector, one needs to modify the GR equations,
as in GR all geometrical terms but $\mathfrak{G}_{\mu\nu}$ are collected in the right hand side of the field equations (\ref{Modified-gravity}). Extensions proposed in order to fulfill this include those built on the basis of Riemannian geometry (curvature based theories), such as  Gauss-Bonnet and $f(R)$ theories \cite{DeFelice:2010aj,Clifton:2011jh,Capozziello:2011et,Nojiri:2010wj}, while others are constructed in the context of  Weitzenb\"{o}ck geometry (torsion based theories); e.g. new general relativity, teleparallel equivalent to general relativity (TEGR) gravity, $f(T)$ theories \cite{Cai:2015emx,Nojiri:2017ncd}. Also, some are constructed in the non-metricity geometry (non-metricity based theories); e.g. symmetric teleparallel equivalent to general relativity (STEGR) \cite{Nester:1998mp} and its recent extension to $f(Q)$ theories \cite{Jarv:2018bgs}.

As mentioned above, both physical and geometrical dark energies could have similar contributions to the field equations (\ref{Modified-gravity}). Nevertheless, they represent fundamentally different physical descriptions; exploration of alternative cosmological models based on modified gravity are thus motivated by theoretical considerations as well as empirical anomalies listed in \cite{DiValentino:2015bja}.

Perhaps the most significant anomaly is embodied in the apparent inconsistency of the locally measured value of the Hubble parameter and that inferred from Cosmic Microwave Background (CMB) observations. It is on this that we focus in this paper, showing how cosmological observations could be made consistent in terms of $f(T)$ theories of gravity with infrared corrections.
The latest released data sets suggest that there is in fact no concordance value for the current Hubble value $H_0$. The local measurements (SNIa and HST) give $H_{0}> 70$ km/s/Mpc \cite{Anderson:2013zyy,Riess:2016jrr,Riess:2018byc,Riess:2018jrr}, on the contrary the global (CMB) measurements  give $H_{0}<70$ km/s/Mpc \cite{Bennett:2012zja,Adam:2015rua,Ade:2015xua}. As the accuracy on both tracks has increased, the tension between these, instead of disappearing,  has crossed over to $3.8$ standard deviations \cite{Riess:2018byc}.
So far no source of systematic uncertainty has been pinpointed to explain the discrepancy of the measurements of the Hubble constant.
This being the case, it  seems natural to investigate new physical inputs, which could restore consistency of the two tracks. However, major changes due to new physics are not supported by the CMB power spectrum.

Possible extensions to the $\Lambda$CDM scenario that have been suggested in order to resolve the aforementioned tension, include  invoking a larger neutrino effective number $N_{eff} \sim 3.5$, i.e. the possibility of a dark radiation component (the standard value is $N_{eff}=3.04$). A second avenue  involves a  phantom dark energy component with an equation of state $w_{DE} \lesssim -1.1$. This could bring the Planck constraint into better agreement with higher values of the Hubble constant. By varying both parameters simultaneously, it has been shown that there is no privilege for dark radiation if allowance is made for dynamical dark energy \cite{DiValentino:2016hlg,DiValentino:2017zyq,Qing-Guo:2016ykt,Zhao:2017cud}. Phantom energy can be shown  to also ameliorate the age conflict \cite{Cepa:2004bc}.

Several analyses have in fact favored such a phantom dark energy scenario (e.g., \cite{Sahni:2008xx,Sahni:2014xx,Qing-Guo:2016ykt,DiValentino:2016hlg,Alam:2016wpf,
DiValentino:2017rcr,DiValentino:2017zyq,Wang:2017yfu,Zhao:2017cud,Dutta:2018vmq}).
Notably, even a viable quintom behavior which allows phantom phase can be achieved without ghost or gradient instabilities, if one extend $k$-essence to kinetic gravity braiding \cite{Deffayet:2010qz}. However, if one insists to work within the GR framework, and assumes the phantom dark energy to be sourced by the matter sector (e.g. ordinary scalar field), ghost instability would not be avoidable due to violation of the dominant energy condition \cite{Carroll:2003st,Carroll:2004hc,Ludwick:2017tox}. This being the case, the choice of the gravity sector as a source of $\mathfrak{T}^{DE}_{\mu\nu}$ is preferable. In this paper, within the frame of the $f(T)$ modified gravity, we argue that the torsional IR correction is a good candidate to source the phantom-like dark energy. Subsequently, it could resolve the current tension in measuring the Hubble constant.

In Section~\ref{Sec:2}, we revisit the teleparallel geometry and briefly discuss  $f(T)$ gravity. In Section~\ref{Sec:3}, we derive the modified Friedmann's equations of the torsional IR correction obtaining the Hubble-redshift relation. In Section~\ref{Sec:4}, we adopt the dynamical system approach showing that the governing equation is as a one-dimensional autonomous system. This allows to analyze its phase portrait and extract some useful information. We show that the model predicts a transitional redshift compatible with observations. Also, we determine the phantom-like nature of the torsional counterpart. Moreover, we find that the model predicts an age of the universe compatible with observations. In Section~\ref{Sec:5}, we fix the model parameters. We show that the torsional IR model reconciles CMB with the local value of $H_0$. In addition, we confront the model with other measured parameters, the electron scattering optical depth $\tau_e$. However, the model is in serious tension with the BAO observations, in particular the angular distance measures. We argue that phantom/phantom-like DE models, in principle, cannot solve the conflict with BAO observations. In Section~\ref{Sec:6}, we conclude the present work. We add Appendix~\ref{App:A}, for some particular values of the model parameters, to give explicit forms of some useful cosmological parameters, time-redshift relation, density parameters and comoving volume element. Also, we show that the scalar fluctuation propagates with a sound speed $0\leq c_s \leq 1$ at all time.
\section{$\lowercase{f}(T)$ Teleparallel Gravity}\label{Sec:2}
In general, one requires the manifold to be differentiable in order to describe dynamical evolution of a physical system under gravity. This can be achieved by defining a compatible differential structure on the manifold. In other words, installing a connection. Let us focus on linear (affine) connections which are used to transport tangent vectors to a manifold between two points along some curve in a covariant way. In modern literature it can be viewed as a differential operator $\tilde{\nabla}$ and known as Koszul Connection,
\[\tilde{\nabla}_{\mu}\partial_\nu:=\Gamma^\alpha{_{\mu\nu}} \partial_\alpha,\]
where $\Gamma^\alpha{_{\mu\nu}}$ are $d^3$ functions ($d$ is the dimension of the manifold) called the connection coefficients of $\tilde{\nabla}$, or simply an affine connection. It is related to the metric by the non-metricity tensor  \cite{Ort2007}
\begin{equation}\label{nonmetricity}
Q_{\mu\nu\rho}:=-\tilde{\nabla}_{\mu}g_{\nu\rho}.
\end{equation}
By taking the combination $\tilde{\nabla}_{\mu}g_{\rho\sigma}+\tilde{\nabla}_{\rho}g_{\sigma\mu}-\tilde{\nabla}_{\mu}g_{\nu\rho}$, one can write a generalized form of the affine connection as
\begin{equation}\label{affine_connection}
   \Gamma_{\mu\nu}{^\rho}=\overcirc{\Gamma}_{\mu\nu}{^\rho}+K_{\mu\nu}{^\rho}+L_{\mu\nu}{^\rho},
\end{equation}
where $\overcirc{\Gamma}{^{\alpha}}{_{\mu\nu}}= \frac{1}{2} g^{\alpha
\sigma}\left(\partial_{\nu}g_{\mu \sigma}+\partial_{\mu}g_{\nu
\sigma}-\partial_{\sigma}g_{\mu \nu}\right)$ is Levi-Civita symmetric connection, $K_{\mu\nu}{^\rho}$ is called the contortion tensor and $L_{\mu\nu}{^\rho}$ is defined in terms of the non-metricity tensor (\ref{nonmetricity}), more geometrical constructions with physical aspects have been reviewed in \cite{Hehl:1994ue}. Notably, the GR has been formulated by requiring a vanishing torsion (contortion) and non-metricity, then all gravitational effects are encoded in terms of the Rimannian curvature of Levi-Civita connection. In the TEGR gravity, it is required to dispel the curvature and the non-metricity which defines Weitzenb\"ock connection, then all gravitational effects are encoded in terms of the torsion tensor of that connection \cite{Maluf:1994mp,Maluf:2013ap}. In STEGR gravity, on the other hand, it is required to have flat connection (null curvature and null torsion tensors), then all gravitational effects are encoded in terms of the non-metricity tensor \cite{Nester:1998mp}. It has been shown that three equivalent variants of GR can be obtained in these three geometries.

In this section, we give a brief description of teleparallel geometry (for more detail see \cite{Aldrovandi:2013wha}) and summarize some of the modifications of the Friedmann equations that can come about in the context $f(T)$ gravity generalization.

In a $4$-dimensional $C^{\infty}$-manifold $(\mathcal{M},\,e_{a})$, where $e_{a}$ ($a=0, 1, 2, 3$) are four linear independent vector (tetrad, vierbein) fields defined on $\mathcal{M}$, the vierbein fields fulfill the conditions $e_{a}{^{\mu}}e^{a}{_{\nu}}=\delta^{\mu}_{\nu}$ and
$e_{a}{^{\mu}}e^{b}{_{\mu}}=\delta^{b}_{a}$, where $\mu=0, 1, 2, 3$ denotes the coordinate components. The Einstein summation convention is applied to both Latin (tangent space coordinates) and Greek (spacetime coordinates) indices.

One can straightforwardly construct the spacetime metric tensor
\begin{equation}\label{metric}
g_{\mu \nu} \equiv \eta_{ab}e^{a}{_{\mu}}e^{b}{_{\nu}},
\end{equation}
where $\eta_{ab}$ is the flat Minkowski metric on the tangent space of $\mathcal{M}$. Consequently, one can define the
Levi-Civita symmetric connection $\overcirc{\Gamma}{^{\alpha}}{_{\mu\nu}}$ and in fact the full machinery of the Riemannian geometry. As can be noticed from (\ref{metric}) that the vierbein has 16 components, while the associated metric has only 10 components which leaves 6 extra degrees of freedom in the vierbein formalism unfixed. On other words, for a given spacetime metric one cannot define a unique vierbein, that is the local Lorentz invariance problem of teleparallel formalism. However, it has been shown that this problem can be alleviated if one allows for flat but nontrivial spin connection \cite{Krssak:2015oua} (see also \cite{Hohmann:2018rwf}).

In the teleparallel geometry one can construct the nonsymmetric (Weitzenb\"{o}ck) linear connection directly from the vierbein\footnote{Remarkably, other linear connections in vierbein space are discussed in detail \cite{NA2007} (for applications, c.f. \cite{1995Ap&SS.228..221M})}, $\Gamma^{\alpha}{_{\mu\nu}}\equiv
e_{a}{^{\alpha}}\partial_{\nu}e^{a}{_{\mu}}=-e^{a}{_{\mu}}\partial_{\nu}e_{a}{^{\alpha}}$, where the vierbein are parallel with respect to this connection
$\nabla_{\nu}e_{a}{^{\mu}}\equiv 0$, and the differential operator $\nabla_{\nu}$ denotes the covariant derivative associated to the Weitzenb\"{o}ck connection. Since $\Gamma^{\alpha}{_{\mu\nu}}$ is nonsymmetric, it defines the torsion tensor $T^\alpha{_{\mu\nu}}\equiv{\Gamma^\alpha}_{\nu\mu}-{\Gamma^\alpha}_{\mu\nu}={e_a}^\alpha\left(\partial_\mu{e^a}_\nu
-\partial_\nu{e^a}_\mu\right)$. However, its curvature vanishes identically. Also, the contortion tensor is given by $K^{\alpha}{_{\mu\nu}} =e_{a}{^{\alpha}}~
\overcirc{\nabla}_{\nu}e^{a}{_{\mu}}$, where the differential operator
$\overcirc{\nabla}_{\nu}$ denotes the covariant derivative associated to the Levi-Civita connection. Notably, the difference of Levi-Civita and Weitzenb\"{o}ck connections defines the contortion tensor of Weitzenb\"ock geometry, $K^{\alpha}{_{\mu\nu}} \equiv \Gamma^{\alpha}_{~\mu\nu} - \overcirc{\Gamma}{^{\alpha}}_{\mu\nu}$, this can be seen directly from (\ref{affine_connection}) in absence of non-metricity. In addition, the torsion and the contortion tensors satisfy the following useful relations $T_{\alpha \mu \nu}=K_{\alpha \mu
\nu}-K_{\alpha \nu \mu}$, while $K_{\alpha
\mu
\nu}=\frac{1}{2}\left(T_{\nu\alpha\mu}+T_{\alpha\mu\nu}-T_{\mu\alpha\nu}\right)$,
where $T_{\mu\nu\sigma} =
g_{\epsilon\mu}\,T^{\epsilon}_{~\nu\sigma}$.

In teleparallel geometry,  the teleparallel torsion scalar
\begin{equation}
T \equiv {T^\alpha}_{\mu \nu}{S_\alpha}^{\mu \nu},\label{Tor_sc}
\end{equation}
is equivalent to the Ricci scalar $R$ up to a total derivative term. In the above, the superpotential tensor ${S_\alpha}^{\mu\nu}$ is defined as
\begin{equation}
{S_\alpha}^{\mu\nu}=\frac{1}{2}\left({K^{\mu\nu}}_\alpha+\delta^\mu_\alpha{T^{\beta\nu}}_\beta-\delta^\nu_\alpha{T^{\beta
\mu}}_\beta\right).\label{superpotential}
\end{equation}
Use the action
\begin{equation}\label{action}
    \mathcal{S}:=\mathcal{S}_{m}+\mathcal{S}_{g}=\int d^{4}x |e|\left(\mathcal{L}_{m}+\mathcal{L}_{g}\right),
\end{equation}
where $|e|=\sqrt{-g}=\det\left({e}_\mu{^a}\right)$. Also we use $\mathcal{S}_{m}$ ($\mathcal{L}_{m}$) and $\mathcal{S}_{g}$ ($\mathcal{L}_{g}$) to represent the actions (Lagrangians) of matter and gravity, respectively. Since the teleparallel torsion scalar (\ref{Tor_sc}) differs from the Ricci scalar $R$ by a total derivative term, the field equations that transpire when using $T$ (in the Einstein-Hilbert action as the gravitation lagrangian) are just equivalent to those with $R$. This is the Teleparallel Equivalent of General Relativity (TEGR) theory of gravity.
\subsection{The matter sector}\label{Sec:2.1}
By varying $\mathcal{S}_{m}$ with respect to the tetrad fields (which has been shown that it is equivalent to vary with respect to the metric \cite{deAndrade:1997ue}), enables one to define the stress-energy tensor of a perfect fluid as
\begin{equation}\label{Tmn}
 \mathfrak{T}_{\mu \nu}=e_{a \mu}\left(-\frac{1}{e}\frac{\delta \mathcal{S}_{m}}{\delta e_{a}{^\nu}}\right)=\rho u_{\mu}u_{\nu}+p (u_{\mu}u_{\nu}+g_{\mu\nu}),
\end{equation}
where $u^{\mu}$ is the 4-velocity unit vector of the fluid.
\subsection{The gravity sector}\label{Sec:2.2}
In the Einstein-Hilbert action, the TEGR has been generalized by replacing $T$ by an arbitrary $f(T)$ function
\cite{Bengochea:2008gz,Linder:2010py,Bamba:2010iw,Bamba:2010wb} similar to the $f(R)$ generalization. The $f(T)$ Lagrangian is
\begin{equation}\label{gravity-Lag}
    \mathcal{L}_{g}=\frac{1}{2\kappa^2}\,f(T).
\end{equation}
By varying the action $\mathcal{S}_{g}$ with respect to the tetrad fields we obtain the tensor
\begin{equation}
\nonumber   \mathfrak{H}_{\mu \nu}=e_{a \mu}\left(\frac{1}{e}\frac{\delta \mathcal{S}_{g}}{\delta e_{a}{^\nu}}\right).
\end{equation}
This gives rise to \cite{Li:2010cg}
\begin{equation}\label{Hmn}
    \mathfrak{H}_{\mu \nu}=\frac{1}{\kappa^2}\left(f_T \mathfrak{G}_{\mu\nu}+\frac{1}{2}g_{\mu\nu}\left(f-Tf_T\right)+f_{TT}S_{\nu\mu\rho}\nabla^{\rho}T\right),
\end{equation}
where $f_{T}$ and $f_{TT}$, stand for $f_{T}=\frac{d f(T)}{d T}$ and $f_{TT}=\frac{d^2 f(T)}{d T^2}$ respectively.
\subsection{The field equations}\label{Sec:2.3}
Using Eqs. (\ref{Tmn}) and (\ref{Hmn}), the variation of the total action (\ref{action}) with respect to the tetrad fields gives the $f(T)$ gravity field equations
\begin{equation}\label{Field-equations}
    \mathfrak{H}_{\mu}{^\nu}= \mathfrak{T}{_{\mu}}{^{\nu}}.
\end{equation}
Equivalently, by substituting from Eq. (\ref{Hmn}), it can be written as
\begin{equation}\label{field_eqns}
\frac{1}{\kappa^2_{eff}} \mathfrak{G}_{\mu\nu}= \mathfrak{T}_{\mu\nu}+\mathfrak{T}^{DE}_{\mu\nu},
\end{equation}
where
\begin{eqnarray}
\nonumber \kappa^2_{eff}&=&\frac{\kappa^2}{f_T}, \\
\mathfrak{T}^{DE}_{\mu\nu}&=&\frac{1}{\kappa^2} \left(\frac{1}{2}g_{\mu\nu}\left(Tf_T-f\right)-f_{TT}S_{\nu\mu\rho}\nabla^{\rho}T\right).\qquad
\end{eqnarray}
It is clear that the general relativistic limit is recovered by setting $f(T)=T$, where $\kappa_{eff}\to \kappa$ and $\mathfrak{T}^{DE}_{\mu\nu}$ vanishes. This allows one to deal with the torsional dark energy on equal footing with the physical one. Although the teleparallel torsion scalar is not local Lorentz invariant, the field equations in the TEGR limit is invariant under local Lorentz transformation (LLT). On the contrary, the field equations of the non-linear $f(T)$ are not in general invariant under LLT \cite{Li:2010cg,Sotiriou:2010mv}. This crucial property makes the $f(T)$ teleparallel gravity different from $f(R)$ gravity. However, it has been shown that a covariant formulation of $f(T)$ gravity can be obtained by including the non-trivial spin connection, see \cite{Krssak:2015oua}, in addition the determination of the spin connection associated to a certain vierbein has been investigated, see \cite{Golovnev:2017dox, Krssak:2017nlv}.
\subsection{$f(T)$ cosmology}\label{Sec:2.4}
We assume that the background geometry of the universe is a flat Friedmann-Lema\^{\i}tre-Robertson-Walker (FLRW). Hence, we take the Cartesian coordinate system ($t;x,y,z$) and the diagonal vierbein
\begin{equation}\label{tetrad}
{e_{\mu}}^{a}=\textmd{diag}\left(1,a(t),a(t),a(t)\right),
\end{equation}
where $a(t)$ is the scale factor of the universe. Using (\ref{metric}) and (\ref{tetrad}), this gives rise to the flat FLRW metric
\begin{equation}\label{FRW-metric}
ds^2=dt^{2}-a(t)^{2}\delta_{ij} dx^{i} dx^{j},
\end{equation}
where the Minkowskian signature is $\eta_{ab}=(+;-,-,-)$. We note that this choice of the vierbein (\ref{tetrad}) leads to consistent field equations without involving any unphysical degrees of freedom for any $f(T)$ theory~\cite{Ferraro:2011us,Krssak:2015oua}. The diagonal vierbein (\ref{tetrad}) directly relates the teleparallel torsion scalar (\ref{Tor_sc})
to Hubble rate as follows,
\begin{equation}\label{TorHubble}
T=-6H(t)^2,
\end{equation}
where $H(t)\equiv \dot{a}/a$ is Hubble parameter, and the ``dot'' denotes
differentiation with respect to the cosmic time $t$. Inserting the
vierbein (\ref{tetrad}) into the field equations (\ref{field_eqns})
for the matter fluid (\ref{Tmn}), the modified
Friedmann equations of the $f(T)$-gravity are,

\begin{eqnarray}
 \frac{3}{\kappa^2} H^2 = \rho+\rho_T&\equiv& \rho_{eff}, \label{FR1T}\\
  -\frac{1}{\kappa^2}\left(3H^2+2\dot{H}\right) = p+p_T&\equiv& p_{eff}, \label{FR2T}
\end{eqnarray}
where $\rho$ and $p$ are respectively the energy density and pressure of the matter
sector, considered to correspond to a perfect fluid. Independently of the above equations, one should choose an equation of state to relate $\rho$ and $p$. Here, we choose the simple linear barotropic case $p=w \rho$, where $w$ is the equation of state parameter. We are interested in evolution during (pressureless) matter domination, we therefore in practice set
$w = 0$. Additionally, the torsional density and pressure in the above equations are
\begin{eqnarray}
\rho_{T} &=& \frac{1}{2\kappa^2}\left[2Tf_T-T-f(T)\right],
\label{Tor-density}\\
p_{T} &=&
\frac{1}{2\kappa^2}\left[\frac{f(T)-Tf_T+2T^2f_{TT}}{f_{T}+2Tf_{TT}}\right],
\label{Tor-press}
\end{eqnarray}
By acquiring the standard matter conservation, we write the continuity equation
\begin{equation}\label{continuity1}
\dot{\rho}+3H(\rho+p)=0.
\end{equation}
This in return implies the continuity equation of the torsional fluid
\begin{equation}\label{continuity2}
\dot{\rho}_T+3H(\rho_T+p_T)=0,
\end{equation}
in order to have a conservative universe. We additionally take an equation-of-state parameter $w_T\equiv p_T/\rho_T$ of the torsional fluid, which incorporates the dark energy sector. So we write
\begin{equation}\label{torsion_EoS}
w_{DE}\equiv w_{T}=\frac{p_{T}}{\rho_{T}}=-1+\frac{[f(T)-2T
f_{T}](f_{T}+2Tf_{TT}-1)}{[f(T)+T-2Tf_{T}]
(f_{T}+2Tf_{TT})}.
\end{equation}
It is useful to define the effective equation-of-state parameter
\begin{equation}\label{eff_EoS0}
w_{eff}\equiv \frac{p_{eff}}{\rho_{eff}}=-1-\frac{2}{3}\frac{\dot{H}}{H^2},
\end{equation}
which can be related to the deceleration parameter $q$ by the following expression
\begin{equation}
q\equiv -1-\frac{\dot{H}}{H^2}=\frac{1}{2}\left(1+3 w_{eff}\right).
\label{deceleration}
\end{equation}
Thanks to the nice feature of the $f(T)$ theory being its field equations second order; currently there are viable $f(T)$ theories of gravity which give good results with a wide range of cosmological observations \cite{Nunes:2016qyp,Bing:2018apj,Nunes:2018xbm}. In the following sections, we focus on a specific model with IR torsional gravity which is in principle an alternative to phantom dark energy.
\section{Torsional IR Correction Model}\label{Sec:3}
In this section, we explore the cosmic evolution that arises as a consequence of the $f(T)$ teleparallel gravity
\begin{equation}\label{fTn}
    f(T)=T+\alpha \frac{~T_{0}^{1+n}}{T^n},
\label{eq:FT}
\end{equation}
where $\alpha$ and $n$ are dimensionless parameters. We denotes the present value of a quantity by a subscript ``0", so $T_0$ is the present value of the teleparallel torsion scalar (using (\ref{TorHubble}), we have $T_0=-6H_0^2$). As a matter of fact, the addition $1/T^n$-term will be effective in the small torsion (i.e Hubble) regime on the large scale, so we would refer to this term as torsional IR correction. As is clear, the model recovers the GR limit at $\alpha=0$ or in the large $T$ regime, where the orders of magnitudes $O(T^n) \gg O(T_0^{1+n})$. On the other hand, it reduces to $\Lambda$CDM at $n=0$ or in the small $T$ regime, as the magnitudes $O(T^{n}) \sim O(T_0^{1+n})$, where the quantity $O(\alpha T_0) \thickapprox O(\Lambda)$. Using relation (\ref{TorHubble}), we write the torsional density and pressure (\ref{Tor-density}) and (\ref{Tor-press}) in terms of the Hubble parameter as
\begin{eqnarray}
  \rho_T &=& \frac{3\alpha}{\kappa^2(2n+1)}H_0^2\left(\frac{H_0}{H}\right)^{2n}, \label{Torsion-density_H}\\
  p_T &=& -\frac{3\alpha}{\kappa^2}\frac{(1+3n+2n^2)H_0^{2n+2}H^2}{H^{2n+2}+\alpha(2n+1)H_0^{2n+2}}. \label{Torsion-pressure_H}
\end{eqnarray}
Inserting (\ref{Torsion-density_H}) into Friedmann equation (\ref{FR1T}) at current time, we write
\begin{equation}\label{alphan}
    \alpha=\frac{1-\Omega_{m,0}}{2n+1},
\end{equation}
where $\Omega_{m,0}=\frac{\rho_0}{3H_0^2/\kappa^2}$ is the current matter density parameter. The above equation shows that only one of the parameters $\alpha$ and $n$ is independent. In addition, we use the constraint $\Omega_{m,0}h^2=0.1417$, where $h=H_0/100$, as estimated by the CMB measurements \cite{Ade:2015xua}, in the rest of our calculations. We will discuss this condition in more details later on.

The continuity equation of the CDM gives $\rho(H)=\rho_0/a(H)^3$, where $\rho_0=3\Omega_{m,0}H_0^2/\kappa^2$ is the current density. Then, the scale factor reads
\begin{equation}\label{scale-factor}
    a^3=\frac{\Omega_{m,0}H_0^2 H^{2n}}{H^{2n+2}-(1-\Omega_{m,0})H_0^{2n+2}}.
\end{equation}
Using the scale factor-redshift relation, $1+z=\frac{a_0}{a}$, where $a_0=1$ at present, we write
\begin{equation}\label{redshift}
    z=\left(\frac{E^{2}-(1-\Omega_{m,0})E^{-2n}}{\Omega_{m,0}}\right)^{\frac{1}{3}}-1,
\end{equation}
where $E=H/H_0$. The inverse relation of (\ref{redshift}) gives $H(z)$, however for a particular $n$ this could be expressed explicitly but complicated. Later in Section \ref{Sec:5}, we show the value $n=1/3$ is preferable by observations. For $n=1$ case, a simpler form of $H(z)$ is given in appendix \ref{App:A} as an example with other features to show the cosmic history according to the torsional IR correction model. So in the following we focus our discussion on the cases, $n=0$ which reduces to the $\Lambda$CDM scenario, in addition to $n=1/3$ and $n=1$ models.
\section{Cosmic history and the phantom regime}\label{Sec:4}
In this section we describe cosmic history in the context of  torsional gravity models with IR corrections of the form described in the previous section. We show these corrections can provide a mechanism for an accelerated phase of cosmic expansion. Prior to this the evolution is essentially equivalent to that of the standard model, thermal history and structure formation are therefore not expected to be affected. We evaluate the transition to the accelerated phase and show that this eventually involves a phantom regime. The measurements  of the deceleration to acceleration  transition and dark energy EoS are not currently precise enough to distinguish our models from the standard one. Current observations that can do are examined in the next section.
\subsection{Phase portrait analysis and deceleration-to-acceleration transition}\label{Sec:4.1}
\begin{figure}
\begin{center}
\includegraphics[scale=0.42]{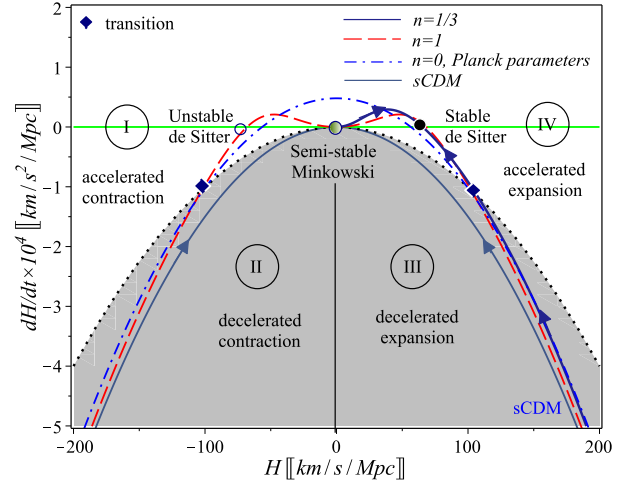}
\caption{\textit{Following the drawing codes in \cite{Awad:2017yod}, the phase portrait (\ref{phase_portrait}) of the torsional IR correction matches the sCDM portrait at $H \gg H_{tr}$, while it intersects the zero acceleration curve at $H_{tr}$ and evolving towards a fixed point $H_f$ values. Thus, the model is in agreement with standard cosmology at past and can perform late acceleration in agreement with observations. We use $w=0$, $\Omega_{m,0}=0.262$ and $H_{0}=73.5$ km/s/Mpc for $n=1/3$ and $n=1$ models, while in $n=0$ model we use Planck parameters.}}
\label{Fig:phase_portrait}
\end{center}
\end{figure}

In a recent study~\cite{Awad:2017yod} (see also \cite{Hohmann:2017jao}), the dynamical system approach was applied to the ordinary differential equations arising in the context of $f(T)$ cosmologies.  This showed that the modified Friedmann equations can be reduced to a \textit{one-dimensional autonomous system}, where $\dot{H}=\mathcal{F}(H)$. This allows to utilize some geometrical procedures to analyze the dynamical behavior of the set of all solutions and its stability just by visualizing it as trajectories in an ($H$, $\dot{H}$) phase space. As seen in Fig.~\ref{Fig:phase_portrait}, the ($H$, $\dot{H}$) phase space has a Minkowskian origin at (0, 0), while by identifying the zero acceleration boundary curve $q\equiv -1-\dot{H}/H^2=0$ (given as dot curve) the phase space is divided into four dynamical regions according to the values of $H$ and $q$ in each region: The unshaded region (I) represents an accelerated contraction, since $H<0$ and $q<0$. The shaded region (II) represents a decelerated contraction, since $H<0$ and $q>0$. The shaded region (III) represents a decelerated expansion, since $H>0$ and $q>0$. The unshaded region (IV) represents an accelerated expansion, since $H>0$ and $q<0$. Notably, one can thus examine and evaluate complicated cosmological models by following their phase trajectories and studying their qualitative behavior, as for example their capability to cross between different regions of the phase space.

It has been proven that the $f(T)$ phase portraits can be analyzed easily and information can be extracted in a clear way (for more applications of this approach to $f(T)$ gravity cosmology see \cite{Bamba:2016gbu,ElHanafy:2017xsm,ElHanafy:2017sih,Awad:2017ign}). In particular, the governing equation is given by
 \begin{equation}\label{phase_portrait_fT}
\dot{H}=3(1+w)\frac{f-H f_{H}}{f_{HH}}= \mathcal{F}(H),
\end{equation}
where $f= f(H)$, $f_H=\frac{df}{dH}$ and $f_{HH}=\frac{d^2 f}{dH^2}$. Inserting the torsional IR correction (\ref{fTn}) into the governing equation (\ref{phase_portrait_fT}), we can determine the phase portrait equation of the model:
\begin{equation}\label{phase_portrait}
    \dot{H}= -\frac{3}{2} (1+w) H^2\left[\frac{(H/H_0)^{2(n+1)}-(1-\Omega_{m,0})}{(H/H_0)^{2(n+1)}+n(1-\Omega_{m,0})}\right].
\end{equation}
At large Hubble regime, the above equation reduces to
\[\dot{H}\thickapprox-\frac{3}{2} (1+w) H^2,\]
which characterizes the phase portrait in general relativity. The torsional IR correction model thus matches
standard predictions  of matter domination ($w=0$), prior to cosmic acceleration, as well as the earlier radiation domination era, $w=1/3$, era. Indeed, from Eqs.~(\ref{Torsion-density_H}) and (\ref{Torsion-pressure_H}), it is not difficult to show that $\rho_T\to 0$ and $p_T\to 0$ as $z\to \infty$ ($H\to \infty$). We thus expect that our torsion correction to the teleparallel equivalent to GR will not affect the thermal history and structure formation up to the transition to cosmic acceleration.

In Fig.~\ref{Fig:phase_portrait}, we visualize the phase portrait (\ref{phase_portrait_fT}) for different values of $n$ verses the $\Lambda$CDM ($n=0$) using Planck parameters. As clear the portrait is unbounded from below, where $\dot{H}\to -\infty$ as $H\to \infty$, which indicates an initial singularity (Big-Bang), asymptotically the portrait matches the sCDM one in the shaded region III (decelerated expansion).  However, it cuts the zero acceleration curve, $q=0$ (i.e. $\dot{H}=-H^2$), which determines the value of the Hubble parameter at transition $H_{tr}$. Using (\ref{phase_portrait_fT}), we find
\begin{equation}\label{Htr}
    H_{tr}=\left[(2n+3)(1-\Omega_{m,0})\right]^{\frac{1}{2(n+1)}}H_{0}.
\end{equation}
Using the values $H_0=73.5$ km/s/Mpc and $\Omega_{m,0}=0.262$, we obtain $H_{tr}=107$ ($102$) km/s/Mpc for $n=1/3$ ($n=1$), respectively. For $n=0$ with Planck parameters ($\Lambda$CDM), we find $H_{tr}=98$ km/s/Mpc. Plugging these results in (\ref{redshift}), we determine the transitional redshift $z_{tr}\sim 0.797$ ($\sim 0.798$) for $n=1/3$ ($n=1$), respectively. However, for $\Lambda$CDM with Planck parameters, it is $z_{tr}\sim 0.649$.

For comparison, we can also pin the predicted transition to cosmic acceleration directly from
the evolution of the deceleration parameter.  Plugging (\ref{phase_portrait}) into (\ref{deceleration}), we write the deceleration parameter
\begin{equation}\label{decelerationn}
    q(z)=-1+\frac{3}{2}\frac{E(z)^{2n+2}-(1-\Omega_{m,0})}{E(z)^{2n+2}+n(1-\Omega_{m,0})}.
\end{equation}
\begin{figure}
\begin{center}
\includegraphics[scale=0.3]{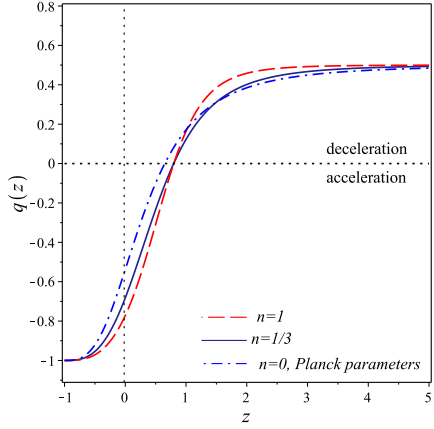}
\caption{\textit{The evolution of the deceleration parameter (\ref{decelerationn}): The plots show that the transition from deceleration to acceleration is at $z_{tr}\sim 0.8$ when $n=1/3$ and $n=1$ which is in agreement with observations. For the torsional IR correction model with $n=1/3$ and $1$, we take $\Omega_{m,0}$=0.26 and $H_{0}$=73.5 km/s/Mpc. For $n=0$ model ($\Lambda$CDM), we use the Planck parameters $\Omega_{m,0}=0.308$ and $H_0=68$ km/s/Mpc.}}
\label{Fig:deceleration}
\end{center}
\end{figure}

In Fig.~\ref{Fig:deceleration}, we plot the evolution of the deceleration parameter for different values of $n$ verses the $\Lambda$CDM using Planck parameters. The plots show that the deceleration parameter $q\to 0.5$ ($w_{eff}\to 0$) at high redshift which is agreement with the sCDM domination. In addition, the transition from deceleration to acceleration occurred at redshift $z_{tr}\thickapprox 0.8$ which is in agreement with the measured value \cite{Farooq:2016zwm}. Also, the current value of the deceleration parameter $q_0\thickapprox -0.68$ ($w_{eff}\thickapprox-0.79$).

The portrait crosses the zero acceleration curve to the unshaded region IV (accelerated expansion) and evolves towards a fixed point $H_{f}$ at $\dot{H}=0$. This determines the Hubble value at the fixed point
\begin{equation}\label{Hfix}
    H_f=(1-\Omega_{m,0})^{\frac{1}{2(n+1)}}H_0.
\end{equation}
Notably, this fixed point cannot be reached in finite time, i.e. $H\to H_f=constant$ as $t\to \infty$, this indicates a pseudo-rip fate \cite{Frampton:2011aa}. In the following we show that this is associated with a phantom regime.

\begin{figure}
\begin{center}
\includegraphics[scale=0.3]{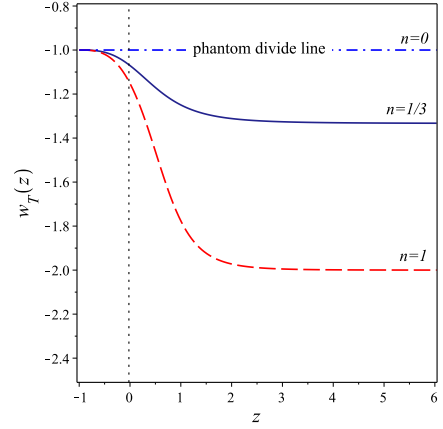}
\caption{\textit{The evolution of the torsional EoS parameter (\ref{EoS_Torsion_H}): For $n=1/3$ and $n=1$, the plots show that the torsional IR correction incorporates a dynamical phantom-like dark energy. At high redshift $w_T\to -1\frac{1}{3}$ (or $-2$) where $n=1/3$ (or $n=1$), respectively. In both models the torsional fluid evolves towards a cosmological constant $w_T\to -1$ at far future, therefore the big-rip fate is avoidable in those models. The current value is $w_T= -1.07$ (or $-1.15$) where $n=1/3$ (or $n=1$) in agreement with observational constraints. For $n=0$, the torsional fluid gives a fixed EoS $w_T=-1$, i.e. cosmological constant.}}
\label{Fig:torsion_EoS}
\end{center}
\end{figure}
\subsection{Phantom-like effective DE}\label{Sec:4.2}
In order to investigate the physics of the torsional IR correction, we define its equation of state (\ref{torsion_EoS}).
Substituting from (\ref{Torsion-density_H}) and (\ref{Torsion-pressure_H}), we obtain
\begin{equation}\label{EoS_Torsion_H}
    w_T(z)\equiv \frac{p_T}{\rho_T}=-1-n\left[\frac{E(z)^{2n+2}-(1-\Omega_{m,0}) }{E(z)^{2n+2}+n(1-\Omega_{m,0})}\right].
\end{equation}
The inverse relation of Eq. (\ref{redshift}) allows to express the torsional EoS in terms of redshift, $w_T(z)$, explicitly. In Fig.~\ref{Fig:torsion_EoS}, we plot the torsional EoS for different choices of the parameter $n$.

We thus determine the current value of the torsional equation of state, $w_T(z=0)= -1.07$ ($-1.15$) where $n=1/3$ ($n=1$), respectively, which is in agreement with observations \cite{Sahni:2014xx,DiValentino:2016hlg,DiValentino:2017zyq}. We find that the torsional fluid at past fixed to $w_T\to -1\frac{1}{3}$ ($-2$) where $n=1/3$ ($n=1$) as the redshift $z\to \infty$, while it is evolving towards the cosmological constant with $w_T\to -1$ at far future as $z\to -1$. This confirms that the torsional IR correction incorporates phantom-like dark energy.

As mentioned in the introduction,  dynamical phantom-like dark energy is in fact favored by recent observations. Modified gravity can provide for a  framework for such scenarios without introducing ghost instabilities associated with scalar field models of phantom dark energy.

Finally, it is worth noting that the invoked phantom regime does not
violate age constraints
 (e.g., such as those  derived from old globular clusters) even while using the locally measure value of $H_0$.
For the proposed model (\ref{fTn}), the age of the universe is
\begin{eqnarray}\label{age}
\nonumber    t_{age}&=&-\int_{H_0}^{\infty}\dot{H}^{-1}~dH\\
                    &=&\frac{2}{3H_0}\int_{1}^{\infty}E^{-2}\frac{E^{2(n+1)}+n(1-\Omega_{m,0})}
                    {E^{2(n+1)}-(1-\Omega_{m,0})}~dE.
\end{eqnarray}
Even for a large Hubble constant, e.g $H_0=73.5$ km/s/Mpc as measured by local observations \cite{Riess:2018byc} (hereinafter referred to as R18) and $\Omega_{m,0}\sim 0.262$ (so as to keep $\Omega_{m,0} h^2=0.1417$ constant as we discuss below), the model predicts an age $t_{age}\sim 13.6$ ($13.9$) billion years for $n=1/3$ ($n=1$). In conclusion, the model predicts an age of the universe compatible with current observations.
\section{Confrontation with observations}\label{Sec:5}
In this section, we fix the free parameters of the torsional IR gravity model, $n$ and $\alpha$ (alternatively $\Omega_{m,0}$). We use Planck measurement of the CMB shift parameter at recombination to constrain the value of $n$ according to the $H_0$ value. In addition, we use the Planck constraint fixing $\Omega_{m,0}h^2\sim 0.1417$ so that we do not have any deviation from the CMB Planck results. Also, we confront the model predictions of the electron-scattering optical depth at reionization with the Planck measurements. Next we use cosmic chronography (CC) and radial and transverse BAO measurements including Lyman-$\alpha$ observations to examine the model.
\subsection{Distance to CMB and shift parameter: resolving the $H_0$ tension}\label{sec:tension}
As is now well known there exists significant tension between the locally measured value of the Hubble constant and that inferred from the CMB. For example, \cite{Riess:2018byc} recently measured $H_0$ = ($73.52  \pm 1.62$ km/s/Mpc), while
\cite{Ade:2015xua} estimate $H_0$ = ($67.8  \pm 0.9$ km/s/Mpc). While the debate continues as to whether the discrepancy is due to new physics or simply observational systematic, it is straightforward to show that the values can in principle be reconciled by invoking a phantom acceleration regime, as we now outline.

Given a primordial fluctuation spectrum and an FLRW cosmology, the relative height of the CMB peaks is essentially determined by the dimensionless physical dark matter and baryon densities $\Omega_{c} h^2$ and $\Omega_{b} h^2$, respectively.
Fixing, in addition,  the number of effective relativistic degrees of freedom  in turn fixes the era of matter radiation equality and recombination, and with these the intrinsic physical scale of the CMB peaks (e.g., \cite{Hu:1994uz}; \cite{Percival:2002gq}), as well as light element production in the context of big bang nucleosynthesis (BBN).  We will assume that all these parameters are fixed to standard values (namely, as quoted in \cite{Ade:2015xua}), and that the cosmological evolution is practically indistinguishable from the standard  scenario up to late times, when the dark energy like component becomes significant. For the specific case of the modified gravity models used here, the latter assumption is justified by the fact that the IR correction theory tends to the teleparallel equivalent to GR at such redshifts, we thus expect the evolution, including the growth of perturbations, to be similar.

In this context, a measurement of the angular diameter (transverse) distance to the CMB
\begin{equation}
D_A (z) =  \frac{1}{1+z}\int_0^{z} \frac{dz'}{H(z')}.\label{ang_dist},
\label{eq:angdd}
\end{equation}
with $z = z_{lss}$ referring to the redshift of last scattering surface,
determines $H_0$,  given a cosmological model (i.e. $H (z)$).
In the standard $\Lambda$CDM model,  such a measurement  should
yield a value that is smaller than locally measured values (similar to the one
obtained by \cite{Ade:2015xua} by fitting the full CMB spectrum).
Nevertheless, as $H(z)$ can be written as $H(z) = H_0 E (z)$, it is easy to see that one can increase
$H_0$, while keeping the angular distance constant, by choosing a model
where $E (z)$ smaller than that associated with $\Lambda$CDM in the redshift range $ 0 \le  z \le z_{lss}$.
This state of affairs  would also  be reflected in the  invariance  of the  ``shift parameter"
of \cite{Efstathiou:1998xx}
\begin{equation}\label{Shift_parameter}
\mathcal{R}_{lss} = \sqrt{\Omega_{m,0}H_0^2}\int_0^{z_{lss}}\frac{dz}{H(z)} =
\sqrt{\Omega_{m,0}}\int_0^{z_{lss}}\frac{dz}{E (z)}.
\label{eq:shift}
\end{equation}
Clearly, if one keeps $\Omega_{m,0} h^2$ constant while increasing
$H_0$, then $\Omega_{m,0}$ becomes smaller.  It is then sufficient for  $E (z)$ to be always below
its  $\Lambda$CDM value $E_\Lambda (z)$ for $0 \le  z \le z_{lss}$, for it to be possible to keep
 $\mathcal{R}_{lss}$ constant.  We now show that this is not only possible
but necessary, in the phantom regime,  that $E_p (z) \le E_\Lambda (z)$.
\begin{figure}
\begin{center}
\includegraphics[scale=0.7]{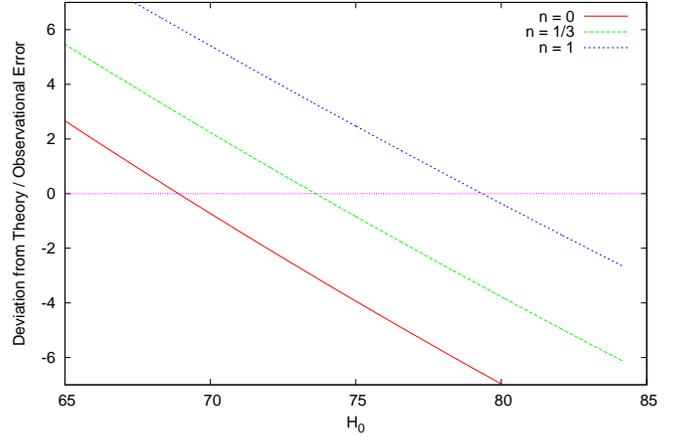}
\caption{\textit{Fixing $H_0$ via CMB shift parameter. The horizontal line of zero error signifies a perfect fit. An index
$n = 0$ corresponds to a cosmological constant, $n = 1/3$ and $n = 1$ refer to departures from this
(given by Eq.~(\ref{eq:FT})).}}
\label{Fig:Shift_parameter}
\end{center}
\end{figure}

Friedmann evolution in a flat universe with matter and DE (or DE-like, as in torsion gravity)
components implies
\begin{equation}\label{E-ratio}
\frac{E_p^2(z)}{E_\Lambda^2  (z)} = \frac{\Omega_{mp} a^{-3} + \Omega_p (z)}{\Omega_{m\Lambda} a^{-3},
+ \Omega_\Lambda}
\end{equation}
where $\Omega_{mp}$ and $\Omega_{m\Lambda}$ refer to the contributions
of the matter densities to the critical density at $z =0$ in the phantom and
$\Lambda$CDM cases respectively. If one requires a larger value for $H_0$
in the phantom case, while keeping $\Omega_{m,0} h^2$ the same in the two cases, then
$\Omega_{mp} < \Omega_{m\Lambda}$. The contribution $\Omega_\Lambda$ to the
current critical density is constant, while $\Omega_p = \Omega_p (z)$, being a phantom DE
contribution, necessarily increases in time (with decreasing $z$).

By definition $E_p  (z =0)/E_\Lambda  (z=0) = 1$. But since $\Omega_p (z=0) >  \Omega_p (z > 0)$, and $\Omega_\Lambda$  is constant, it follows that $\frac{E_p (z)}{E_\Lambda  (z)} < 1$
for $z > 0$, even if $\Omega_{mp} = \Omega_{m\Lambda}$. If we require that
$\Omega_{mp} <  \Omega_{m\Lambda}$, so as to keep $\Omega_{m,0} h^2$ the same while increasing
$H_0$, then the ratio $\frac{E_p  (z)}{E_\Lambda  (z)}$ becomes smaller still.
It is thus apparent that in the presence of phantom like dark energy,
it is not only possible but necessary to decrease $E(z)$
relative to the standard case, which in turn necessitates  an
increase in $H_0$  if CMB angular distance, shift parameter and physical matter densities are
to be kept constant.

Fig.~\ref{Fig:Shift_parameter} illustrates this in the context of our torsion gravity models. Here we vary $H_0$, keeping $\Omega_{m,0} h^2 = 0.1417$  (as measured in \cite{Ade:2015xua}) , and evaluate the shift parameter $\mathcal{R}_{lss}$, substituting $H(z)$, namely the inverse of (\ref{redshift}), into (\ref{eq:shift}) where $z_{lss}=1089.9$ \cite{Ade:2015xua}.  We then subtract this from the measured value of $\mathcal{R}_{lss}=1.7488$, retrieved from (Planck TT+lowP) \cite{Ade:2015rim}, and divide by the error estimate quoted therein ($\pm 0.0074$). As can be seen, as one deviates from the cosmological constant scenario ($n = 0$) and further into the phantom regime, the lines intersect the zero error horizontal at larger values of $H_0$; as, expected, these larger values are thus necessary in order to fit the CMB data embodied in the shift parameter.  In particular a value of about $n = 1/3$ fits the shift parameter with $H_0 = 73.5$ km/s/Mpc, as locally measured by \cite{Riess:2018byc}.

\subsection{Reionization redshift}\label{Sec:5.2}
The electron-scattering optical depth $\tau_e$ of the CMB, provides a direct probe of the reionization epoch and its redshift $z_{re}$; it places constraints on the cosmological model, as it depends on $H (z)$  at redshifts intermediate between $z_{lss}$ and local measurements. The optical depth can be evaluated from
\begin{equation}
\tau_e(z_{re}) = c \int_0^{z_{re}}\frac{n_e(z)\sigma_T \, dz}{(1+z)H(z)} \, , \label{optical_depth}
\end{equation}
where $n_e(z)$ is the electron density and $\sigma_T$ is the Thomson cross-section describing scattering between electrons and CMB photons. Here we take the densities of hydrogen, helium and electrons, respectively, as $n_{H}=\left[(1-Y_p) \Omega_b \rho_{cr,0}/ m_H \right](1+z)^3$, $n_{He}=y n_{H}$ and $n_e=(1+y)n_{H}$, where $y=\frac{Y_p}{4(1-Y_p)}$ and $m_H$ is the hydrogen mass  \cite{Shull:2008su}. We use the Planck constraint $\Omega_{b,0} h^2=0.02230$ \cite{Ade:2015xua}, which gives the baryon density parameter $\Omega_{b,0}=0.0413$ for $H_0=73.5$ km/s/Mpc, the helium mass fraction $Y_p=0.247$ \cite{Peimbert:2007vm} and the current critical density $\rho_{c,0}=1.88\times 10^{-29}\, h^2$ g/cm$^3$. Then, using the inverse function of Eq. (\ref{redshift}) and by evaluating the integral (\ref{optical_depth}), we get $\tau_e(z_{re})\thickapprox 0.058$ at $z_{re} = 8.1$, which is in agreement with \cite{Ade:2015c} (lollipop + PlanckTT + lensing) observations\footnote{We note that torsional IR gravity model is in excellent agrement with the latest Planck results \cite{Aghanim:2018eyx}. Using $\Omega_{b,0} h^2=0.02242$ which gives the baryon density parameter $\Omega_{b,0}=0.0415$ for $H_0=73.5$ km/s/Mpc, $Y_p=0.2454$ as predicted by BBN, we get $\tau_e(z_{re})\thickapprox 0.0553$ at $z_{re} = 7.82$, which is in agreement with \cite{Aghanim:2018eyx} observations (TT,TE,EE+lowE+lensing+BAO), $\tau_e(z_{re})=0.0561\pm 0.0071$ where $z_{re}=7.82 \pm 0.71$.}, $\tau_e(z_{re})=0.058\pm 0.012$ where $7.8<z_{re}<8.8$.

\subsection{Local Hubble parameter evolution}\label{Sec:5.3}
\begin{figure}
\begin{center}
\includegraphics[scale=0.47]{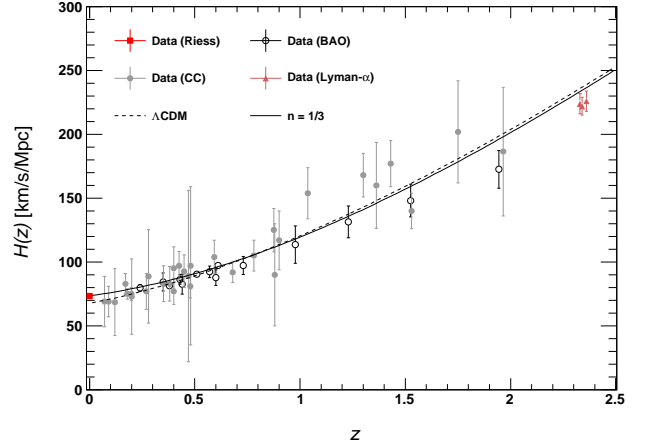}
\caption{\textit{Evolution of Hubble function in terms of redshift: For the $\Lambda$CDM model with $n=0$, we take $H_0=68$ km/s/Mpc. For the torsional IR gravity model with $n=1/3$, we take $H_0=73.5$ km/s/Mpc.}}
\label{Fig:Hubble}
\end{center}
\end{figure}

As the resolution of the $H_0$ tension in terms of phantom dark energy described
above involves changing the evolution of $H (z)$ --- through changing $E (z)$ ---
it is natural to inquire whether this change
can be actually distinguished directly from local $H(z)$ measurements.
Fig.~\ref{Fig:Hubble} collects such measurements.  These include the 43 Hubble measurements given in~\cite{Cao:2017gfv},
which lists  a number of CC and BAO measurements
(including two Ly-$\alpha$ observations).
In addition to the following four BAO measurements: $H(z=0.978)=113.72\pm 14.63$, $H(z=1.23)=131.44\pm 12.42$, $H(z=1.526)=148.11\pm 12.75$ and $H(z=1.944)=172.63\pm 14.79$ km/s/Mpc (\cite{Zhao:2018jxv}), and one BAO Ly-$\alpha$ observation $H(z=2.33)=224\pm 8$ km/s/Mpc  (\cite{Bourboux:2017cbm}). This, in addition to the R18 observation of $H_0=73.52\pm 1.62$ km/s/Mpc as measured by \cite{Riess:2018byc}. Fig.~\ref{Fig:Hubble} shows clearly the capability of the torsional IR gravity (with $n=1/3$) to fit with R18 and Ly-$\alpha$ better than $\Lambda$CDM model.

\begin{table}
\caption{the $\chi^2$ calculations of $H(z)$: For $n= 0$ ($\Lambda$CDM), we take $H_0=68$~km/s/Mpc. For $n=1/3$ (torsional IR gravity), we take $H_0=73.5$~km/s/Mpc.}\label{Table:X2}
\begin{center}
\begin{tabular}{lcccc}
\hline\hline
Dataset                & $n$   & $\chi^2$ / dof &         & $\chi^2_{\text{dof}}$\\
\hline
CC                     & $0$   & 14.77 / 29 &  $\approx$  & 0.51 \\
                       & $1/3$ & 16.78 / 29 &  $\approx$  & 0.58 \\
\hline
BAO                    & $0$   & ~~9.59 / 12 &  $\approx$   & 0.80 \\
                       & $1/3$ & 11.39 / 12 &  $\approx$  & 0.95 \\
\hline
CC+BAO+R18             & $0$   &  35.97 / 17 &  $\approx$ & 0.82 \\
                       & $1/3$ &  28.17 / 17 &  $\approx$ & 0.64 \\
\hline
CC+BAO+R18+Ly-$\alpha$ & $0$   & 44.98 / 47 &  $\approx$  & 0.96 \\
                       & $1/3$ & 33.67 / 47 &  $\approx$  & 0.72 \\
\hline\hline
\end{tabular}
\end{center}
{Note: For R18, we take $H_0=73.52\pm 1.62$ km/s/Mpc as measured by Milky Way 50 Gaia + HST, Long P Parallaxes at redshift \cite{Riess:2018byc}.}
\end{table}

The $\chi^2$ statistics of these forty nine Hubble measurements are
\begin{equation}\label{Chi_Hubble}
    \chi^2(n,\alpha)=\sum_i \frac{\left(H_i^t-H_i^o\right)^2}{\sigma_{H^o_i}^2},
\end{equation}
where the subscript $i=1,2,\ldots,49$, the superscripts $t$ and $o$ denote respectively the theoretical and the observed values of $H(z_i)$, and $\sigma_{H^o_i}$ denote the one standard deviations in the measured values. As can be inferred from Table~ \ref{Table:X2}, both  $\Lambda$CDM and the model with $n = 1/3$
and $H_0 = 73.5$ km/s/Mpc, display $\chi^2/\text{dof} \la 1$, where 'dof' is the number of degrees of freedom given that there are two model parameters. But  the model associated with phantom-like effective
dark energy component  performs better than that invoking cosmological constant.
This remains the case as long the Ly-$\alpha$ and R18 data are included.
The BAO data on its own, on the other hand,  favors the standard model.
As we see below, this  conclusion is definitely consolidated by BAO distance measurements,
combined with the CMB.

We note that the Hubble function is related to the luminosity-distance $D_L(z)$ by
\begin{equation}\label{Hubble_lum_dist}
    H(z)=\left[\frac{d}{dz}\left(\frac{D_{L}(z)}{1+z}\right)\right]^{-1}.
\end{equation}
Since the Hubble function is related to the first derivative of $D_L$, one expects the measured values of $H(z)$ to be much noisier than $D_L(z)$ measurements. On other words, distances are in principle integrable quantities, which makes them relatively more precise. In the following we confront the torsional IR gravity with the BAO angular distance measurements.
\subsection{BAO distance measurements}\label{Sec:5.4}
BAO can be used as standard rulers; from isotropic measurements one can infer $D_V =  \left[cz(1+z)^2 D^2_A(z)/H(z)\right]^{1/3}$
and from anisotropic measurements $D_A$ itself (given
the sound horizon at the baryon drag epoch $r_d$).
These measurements rely on the same principle as that used to infer
the angular diameter distance to the CMB (as once the physical densities
and eras of recombination and matter radiation equality are determined,
the physical scale of the peaks is fixed).
It turns out that such measurements are highly constraining
and essentially rule out solutions of the $H_0$ tension invoking
phantom-like dark energy.

\begin{figure*}[t]
\gridline{\fig{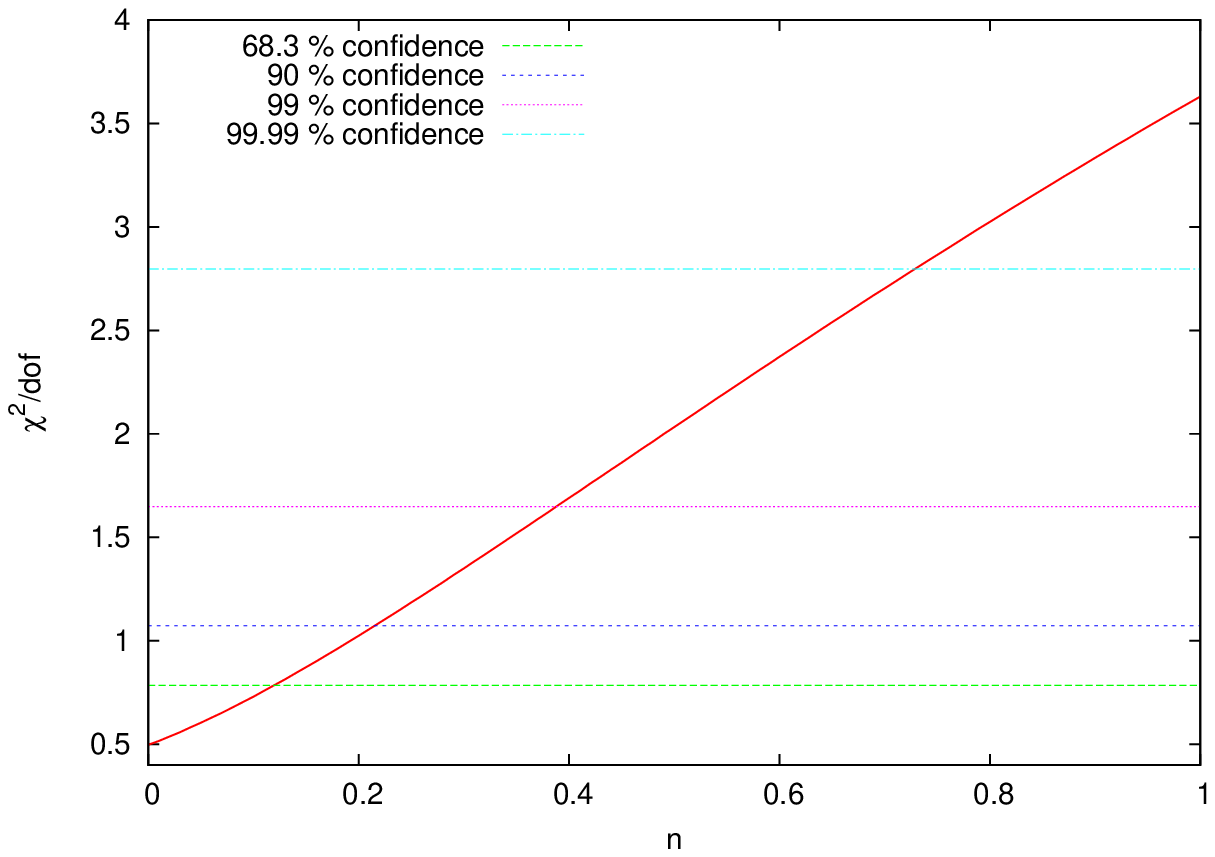}{0.45\textwidth}{}
          \fig{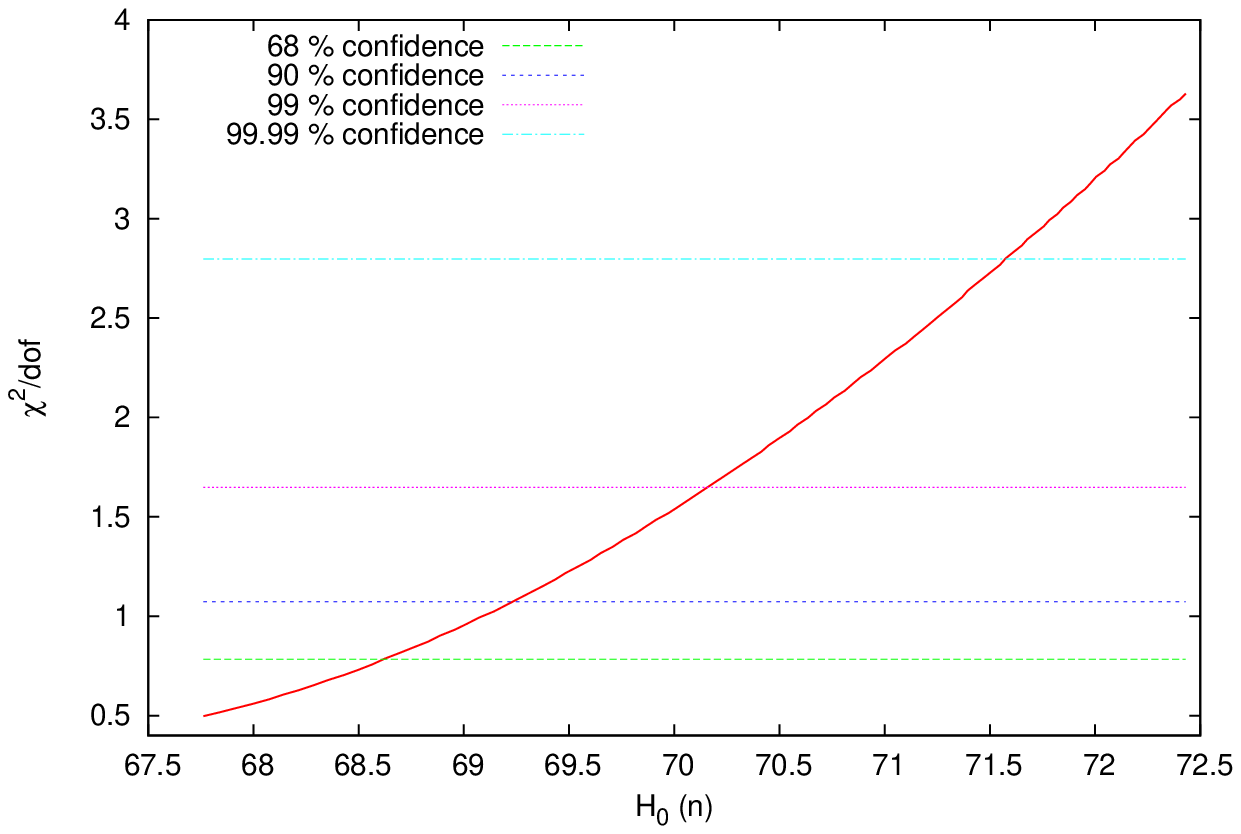}{0.45\textwidth}{}
          }
\caption{\textit{Combined $\chi^2$ per degree of freedom of CMB and BAO distances., with horizontal lines
showing the  associated confidence levels. Left panel: as a function of $n$ (measuring deviation
from cosmological constant at $n = 0$). Right panel: in terms of the value of $H_0$
that minimizes $\chi^2$ for each value of $n$ (assuming $\Omega_{m 0} h^2 = 0.1417$).}}
\label{Fig:Fixing_n_H}
\end{figure*}

We show this by using the observations from various independent data sets:
\cite{Beutler:2011MNRAS}  which use 6dFGS data,
\cite{Kazin:2014qga} for reconstructed WiggleZ,
\cite{Ross:2014qpa}
for the SDSS MGS data, \cite{Alam:2016hwk} for BOSS, , \cite{Ata:2017dya} for eBOSS quasar data.
and \cite{Abbott:2017wcz} for the DES survey.
We use $r_d = 147.5$ Mpc, when the results are given as ratios involving $r_d$.
We calculate the relevant distances  to the observed redshift of the BAO peaks of each observation for our torsion models,
deriving $D_A$ and $D_V$ for values of $0 \le n \le 1$. For each such value, we vary $\Omega_m$ to search for the minimum of
\begin{equation}
\chi^2 (n, H_0)  = \sum_i \frac{(D^t_i - D^o_i)^2}{\sigma_i^2} +
\frac{(\mathcal{R}^t_{lss} - \mathcal{R}^o_{lss})^2}{\sigma^2_{lss}},
\label{eq:Chai}
\end{equation}
where $D_i$ refers to the different BAO measurements (either $D_V$ or $D_A$, depending
on the particular set of observation),
$\sigma$ the one standard deviations
in measurements, and the superscripts $t$ and $o$ refer to the
theoretical and observed values of the different quantities.
For each $n$ there is then a unique $\Omega_m$ that minimizes
the $\chi^2$. Assuming, as we do, that $\Omega_{m,0} h^2$ is held fixed
(at $0.1417$), one can also associate a unique $H_0$ with each $\Omega_m$,
and hence for each $\Omega_m$ that minimizes $\chi^2$ at each $n$.

The results are shown in Fig~\ref{Fig:Fixing_n_H}.
As can be seen, the deviation between the observed and inferred
distances, as measured using~(\ref{eq:Chai})
is smallest for $n=0$. Values of $n \ga 1/5$ are ruled out at the $99 \%$ confidence level.
Moreover for  $n = 1/3$, the corresponding $H_0$ that minimizes $\chi^2$
is significantly smaller than that inferred when fitting the CMB alone in
Section~\ref{sec:tension} above. The reason for this failure is
discussed in the next section.
\subsection{BAO distance measurements and the failure of phantom models}\label{Sec:5.6}
In Section~\ref{sec:tension}, we argued that the angular diameter distance to the CMB and the shift parameter can be kept constant if one increases the value of $H_0$ while invoking cosmic evolution in the phantom regime. This was because $E (z) = H(z)/H_0$  is smaller up to $z=0$ for such scenarios than in the case when the DE contribution comes from a cosmological constant. Requiring a larger value of $H_0$ evidently implies that the $H(z)$ associated with phantom dark energy becomes larger than that of $\Lambda$CDM at some redshift $z_c \ge 0$. From the Friedmann equations
\begin{equation}
\frac{H^2_p (z)}{H^2_\Lambda (z)} =  \frac{\rho_{mp} (z) + \rho_p (z)}{\rho_{m\Lambda} (z)
+ \rho_\Lambda},
\label{eq:ratH}
\end{equation}

\begin{figure*}[ht!]
\gridline{\fig{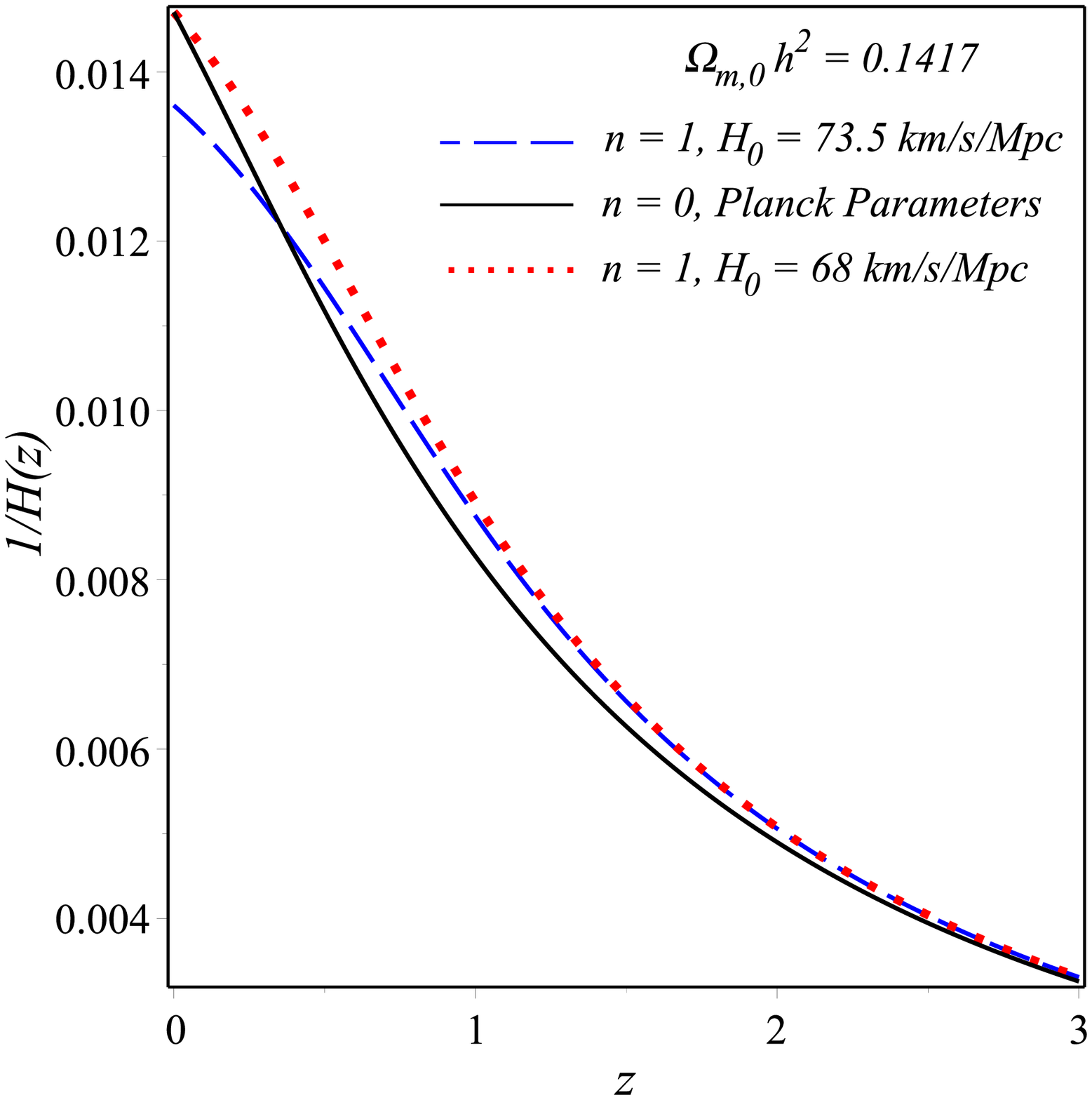}{0.38\textwidth}{}
          \fig{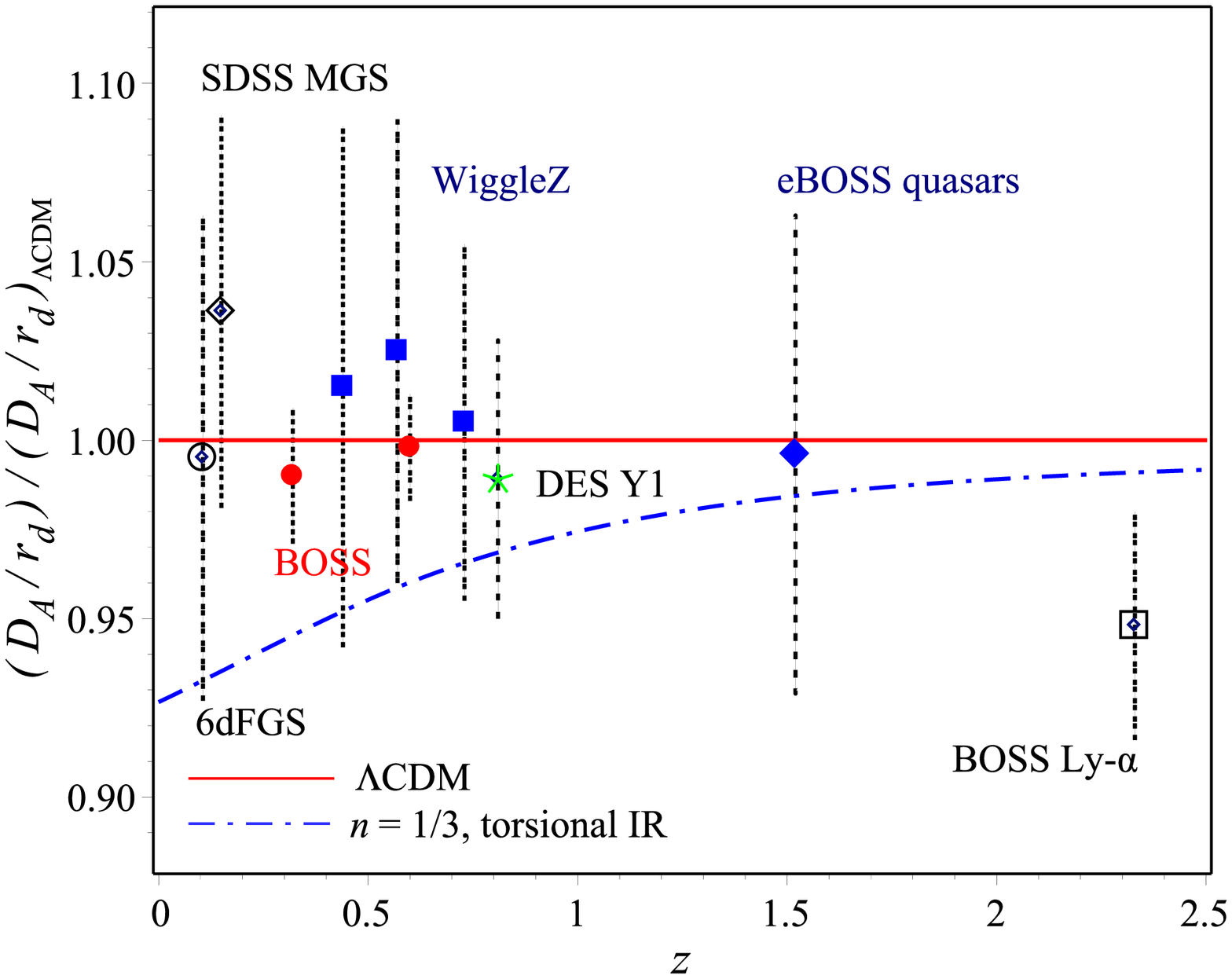}{0.48\textwidth}{}
          }
\caption{\textit{The evolution of the radial and angular distances. Left panel: The redshift evolution of the integrand in Eq.~(\ref{eq:angdd}).
Note that for $n\neq 0$
--- i.e., deviation from a cosmological constant and into the phantom regime ---
$1/H (z)$ is invariably larger if $H_0$ is kept fixed. This implies larger distances for all $z$.
For larger $H_0$ the curves of $n = 0$ and $n = 1$ cross. This means that radial distances
can be underestimated or overestimated, depending on their location relative to the
crossing point. However if the distance to the CMB is fixed, in both models, then the
$n >0$ distances are again invariably smaller. Right panel: Comparison of $\Lambda$CDM (horizontal line)
and phantom model with $n = 1/3$ and $H_0 = 73.5$ km/s/Mpc with BAO transverse distance estimates. As can be seen phantom models systematically underestimate the transverse distance compared to the $\Lambda$CDM, while no such systematics in the data at redshift $z\lesssim 1.5$.}
 \label{Fig:No_systematics}}
\end{figure*}

\begin{figure*}
\gridline{\fig{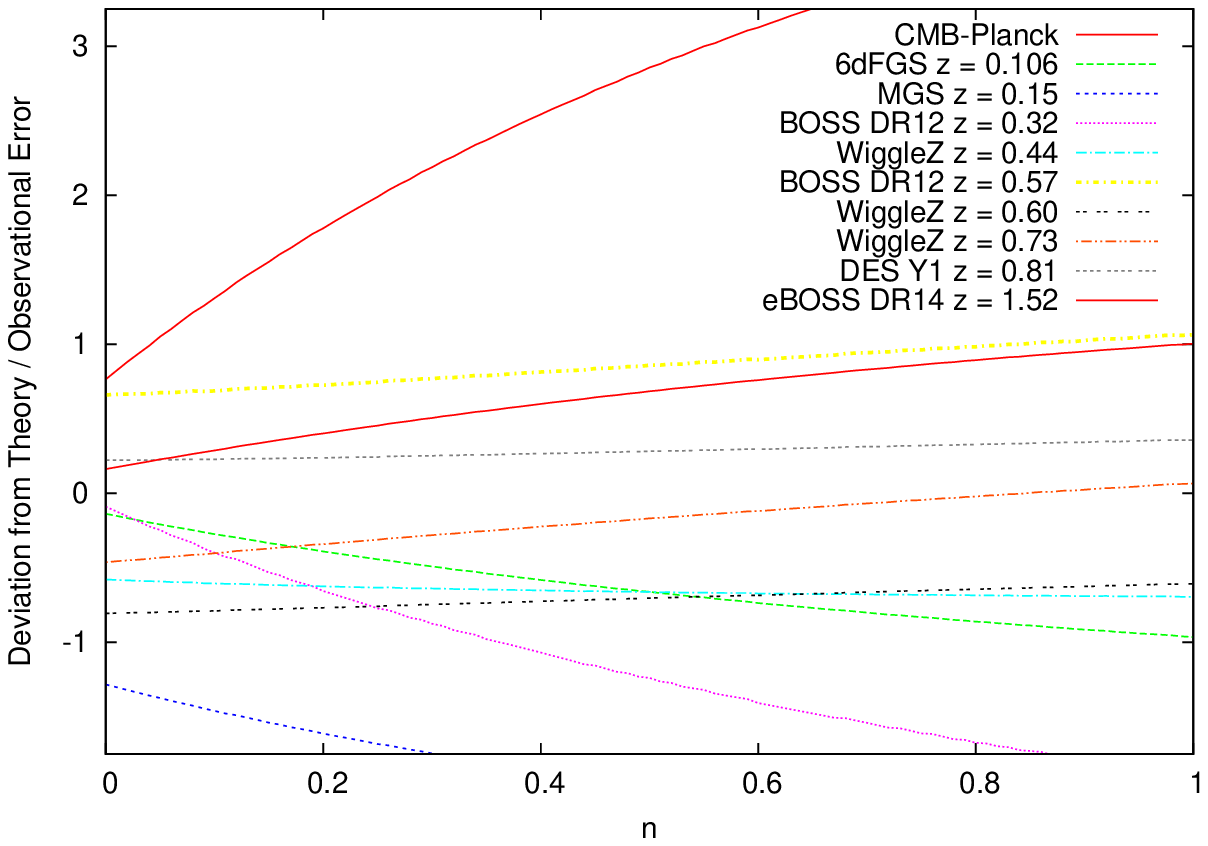}{0.45\textwidth}{}
          \fig{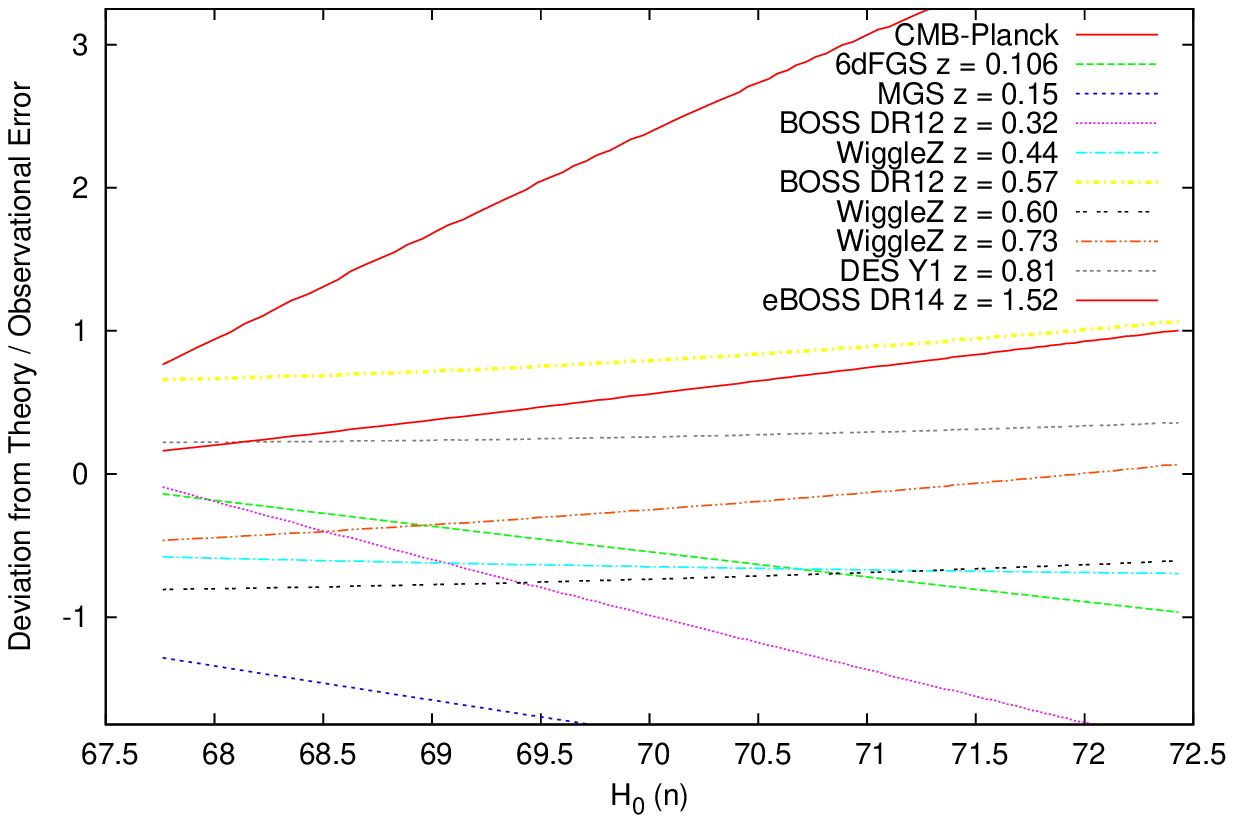}{0.45\textwidth}{}}
\caption{\textit{Same as in Fig.~\ref{Fig:Fixing_n_H}, but for the individual components of the error vector, showing the error evolution in each observable, with change in  $n$, and associated $H_0$.}}
\label{Fig:Errors_n_H}
\end{figure*}

one can find this redshift.
The physical matter densities remain such that $\rho_{mp} = \rho_{m\Lambda}$  if we keep
$\Omega_{m,0} h^2$ fixed, so that the ratio in (\ref{eq:ratH}) is smaller than one when
when the phantom dark energy density is less than that associated with a cosmological constant:
$\rho_p (z) <  \rho_\Lambda$. The ratio then increases to finally reach $H^2_p (z = 0)/ H^2_\Lambda (z = 0)$ at $z=0$, which is greater than unity
 if we assume a larger value of the Hubble constant to be associated with the phantom case.
The critical value
$z_c$ corresponds to a ratio one. If this occurs during matter domination, then the epoch
where  $\frac{H^2_p (z)}{H^2_\Lambda (z)} < 1$ has negligible effect on the evolution and
${H^2_p (z)} > {H^2_\Lambda (z)}$ for all practical purposes (that is during DE domination).
In this case, the angular diameter distance to the CMB will increase. If the this distance (and
shift parameter) are to be kept in line with observed values then
${H^2_p (z)} > {H^2_\Lambda (z)}$
should become unity at $z_c \sim 1$.

In the case of torsion gravity models studied here this is illustrated in Fig.~\ref{Fig:No_systematics},
where we plot $1/H(z)$ --- the integrand in the formula for the angular diameter distance ---
 for the standard model with $n = 0$ and for $n=1$. If $H_0$ assume to be  the same in the two case
$H (n= 0)$ is always smaller or equal to $H (z =1)$, which simply reflects the fact that $E_p \le E_\Lambda$ up to $z=0$ as expected from  the discussion following Eq. (\ref{E-ratio}). When $H_0$ associated with the $n=1$ case is larger the lines cross at $z = z_c$. What this implies is that the angular diameter distance will be smaller than in the standard case for objects at $z < z_c$. And if the distance to the CMB $D_A (z = z_{lss})$ is to be kept fixed, while increasing $H_0$ and invoking the phantom regime, then $D_A (z)$ to any object at $0  \le z \le z_{lss}$ will be larger or equal to that predicted by
$\Lambda$CDM. If the distance to the CMB is overestimated, then the distances to objects can be either overestimated or underestimated depending on its redshift.

This implies that, in order to fit CMB and BAO distances simultaneously using a larger $H_0$ and $n > 0$, the standard model should systematically overestimate distances to BAO measurements, with the discrepancy being maximal for redshifts around $z_c$.
This is not observed, as can be seen from Fig.~\ref{Fig:No_systematics} (right panel). To further illustrate the point, we plot the errors associated with the different observations, which were used to estimate the $\chi^2$ in Fig.~\ref{Fig:Errors_n_H}. As can be seen, at $n = 0$  some distances are overestimated and some underestimated, with no clear trend in terms of redshift dependence. As $n$ is varied, the critical redshift $z_c$ changes, and the $\chi^2$ minimization procedure causes the distance to the CMB to also shift. As a result there is another critical redshift below which BAO measurements are underestimated relative to standard case, and beyond which they are overestimated. Since there is no systematic deviation with respect to $\Lambda$ CDM predictions in the BAO data used, this process means that some distances that were initially underestimated at $n = 0$ become even more so for $n > 0$, and conversely some  overestimates
are increased.

Current CMB and BAO measurements seem to therefore rule out significant phantom-like regime in the redshift range of the BAO data included here. This is the case even if one keep $H_0$ at a small value; for this would  shift the distance to the CMB and also the BOA points due to the smaller $E(z)$ and hence larger associated $1/ H(z)$ (as discussed in Section~\ref{sec:tension} and reflected in Fig.~\ref{Fig:No_systematics}. We note nevertheless that there seems to be a systematic  underestimate of the BAO distances inferred from Lyman-$\alpha$ measurements in the context of $\Lambda$CDM. We have not included these points here, as they lead to worse $\Lambda$CDM fits and do not lead to much improvement for the cases with $n \ne 0$, given that the models studied here are close to $\Lambda$CDM for the relevant redshifts ($z \ga 2$) and the relatively  the large observational errors. Possible modest phantom evolution confined to redshifts  $z \ga 2$ are therefore not ruled
out and can be tested by upcoming data.

\section{Conclusion}\label{Sec:6}

The results presented here suggest  that the torsional IR corrections to teleparallel gravity
lead to a phantom-like effective dark energy term in the Friedmann equations.
Given the current matter density our family of models contain only one free parameter.
A phantom-like dark energy evolution, sourced by the gravitational sector can be derived for
positive values of this parameter  without invoking a canonical scalar field that suffers from ghost instabilities.  We perform a dynamical
system analysis that elucidates the basic qualitative evolution of the system, including the
transition to the accelerated regime.

As has recently been noted, the phantom regime provides a basis for resolving the tension between local and global measurements of the Hubble constant $H_0$. We find that these can indeed be reconciled by our model. For values of the parameter that completely reconcile the two values, the phantom regime comes with  a dynamical equation of state $-1\frac{1}{3}\leq w_{T}(z)\leq -1$ with $w_T= -1.07$ at present. These corresponding deceleration parameter $q_{0} = -0.68$ and effective equation of state $w_{eff,0}=-0.79$ at present, with transition redshift $z_{tr}\approx 0.8$. The model also predicts an electron scattering optical depth $\tau_e\thickapprox 0.058$ at reionization redshift $z_{re} \sim 8.1$,  which is in agreement with observations.

The model however faces serious problems when baryon acoustic oscillation data are included. This is true for both line of sight measurements, from which the Hubble parameter can be inferred and transverse ones yielding measures of the distances to the BAO peaks at different redshifts. The latter case being most severe; with the model parameter $n$ that corresponds to the reconciliation of the local and CMB values of is ruled out to more $99.99  \%$ confidence by these data.

We argue that this failure should be  a generic feature of phantom dark energy models, particularly ones that may solve the $H_0$ tension by predicting currently observable deviations from $\Lambda$CDM evolution at $z \la 2$. For, assuming that $\Omega_{m,0} h^2$ is held constant, so as not to modify the heights of the CMB acoustic peaks, one finds that in fact distances to objects in whole redshift range to the  CMB last scattering surface  are necessarily overestimated, if the angular diameter distance, and associated shift parameter, are to be kept fixed to current observations. Therefore, if  the model   predicts currently observable deviations from $\Lambda$CDM evolution at $z \la 2$,  then it necessarily contradicts the BAO measurements at these redshifts, which do not show any such systematic discrepancies with the standard model. If the distance to the CMB is allowed to shift then the distance to some objects (beyond some critical redshift) will be underestimated and some (at lower redshift) underestimated, again in a systematic way that is not in line with observations. In this case, we mention some scenarios that possibly resolve the conflict with the angular distance measurements: (i) Phantom models with a sudden ripping behavior at low redshift. As see from Fig. \ref{Fig:No_systematics} (the right panel), the non-systematics of the data in fact fit well with models similar to $\Lambda$CDM at low redshifts $z \la 1.5$, however in order to fit with large $H_0$ the model needs to suddenly evolve to phantom regime at $z \la 0.07$, such models may evolves to big-rip singularity or at the best scenario towards a pseudo-rip. In the later one should calculate the ripping inertial force. (ii) Oscillating DE models with quintom behavior (i.e. oscillating about $\Lambda$CDM), where phantom behavior should show up at law redshifts $0 < z \la 0.1$ and $z \la 1.5$, quintessence behavior at an intermediate region $0.1 \la z \la 1.5$, and matches $\Lambda$CDM at larger redshifts. (iii) Non-flat models, where the contribution of the curvature density parameter $\Omega_k$ to the angular distance could provide a correction for better matching with the measured values.

We note that Lyman-$\alpha$ BAO observations at $ z \ga 2$ do indeed currently suggest a systematic underestimate on the part of the standard $\Lambda$CDM of the distances involved. If these persist with incoming measurements, they  could in principle be explained by a phantom regime confined to a range around that redshift.


\acknowledgments

We would like to thank Adi Nusser and Joe Silk for helpful communication.
This work was supported by grant number 25859 from the Egyptian Science and Technology Development
Fund Basic and Applied Research Grants.
\appendix
\twocolumngrid
\section{Example: $\lowercase{n}=1$ Case}\label{App:A}
As mentioned earlier that Eq. (\ref{redshift}) can be inverted to give an explicit Hubble-redshift relation for a particular choice of $n$. However, this form is complicated to be given in detail. For $n=1$ model, the formulae are not complicated and can be given explicitly. Since qualitative features are similar to those are discussed for smaller $n$ values, we present $n=1$ model in detail. In addition, we examine the torsional IR gravity on the perturbation level of the theory by investigating the sound speed $c_s$ of the scalar fluctuations.
\begin{figure*}[th!]
\gridline{\fig{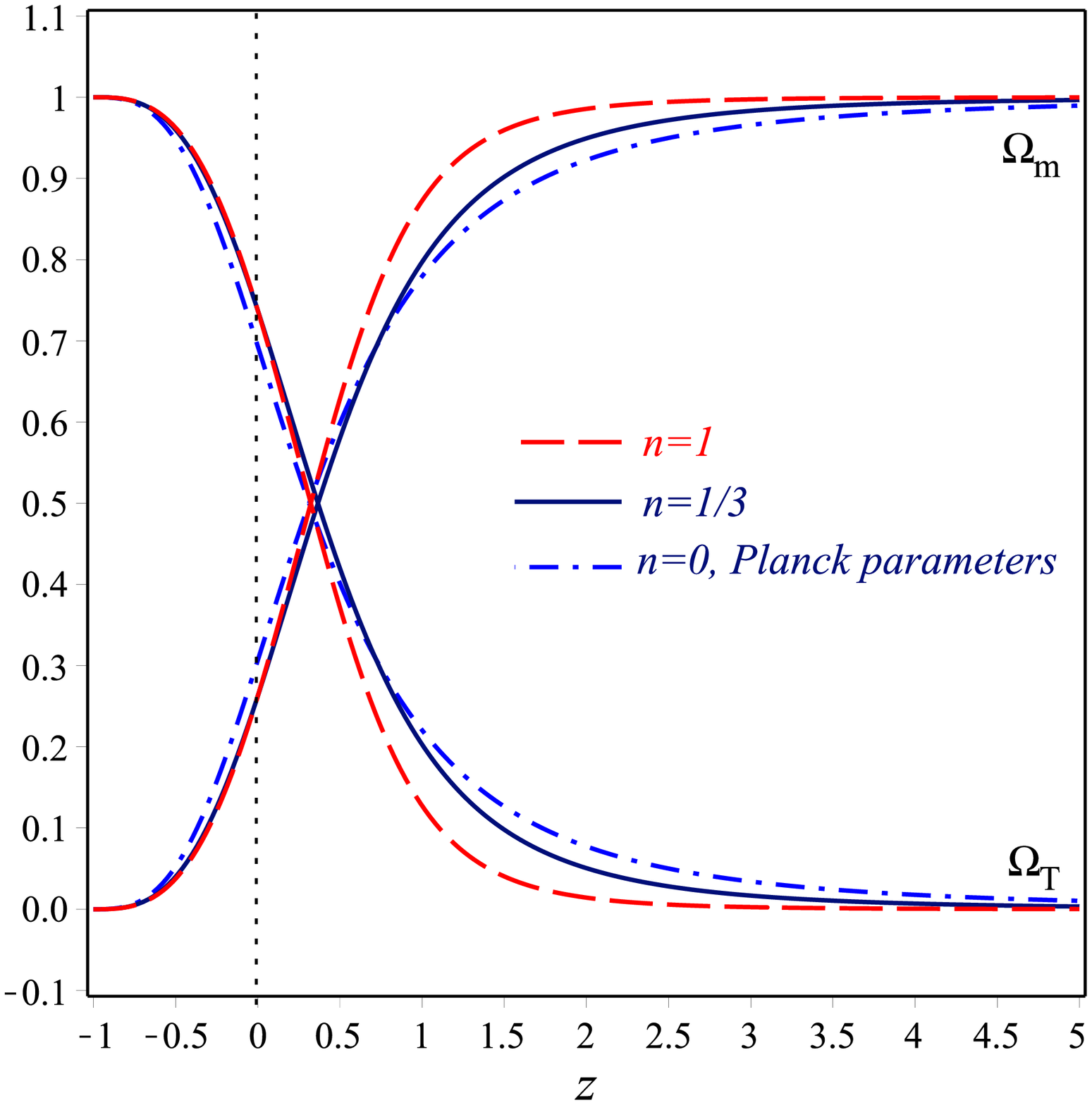}{0.3\textwidth}{}
          \fig{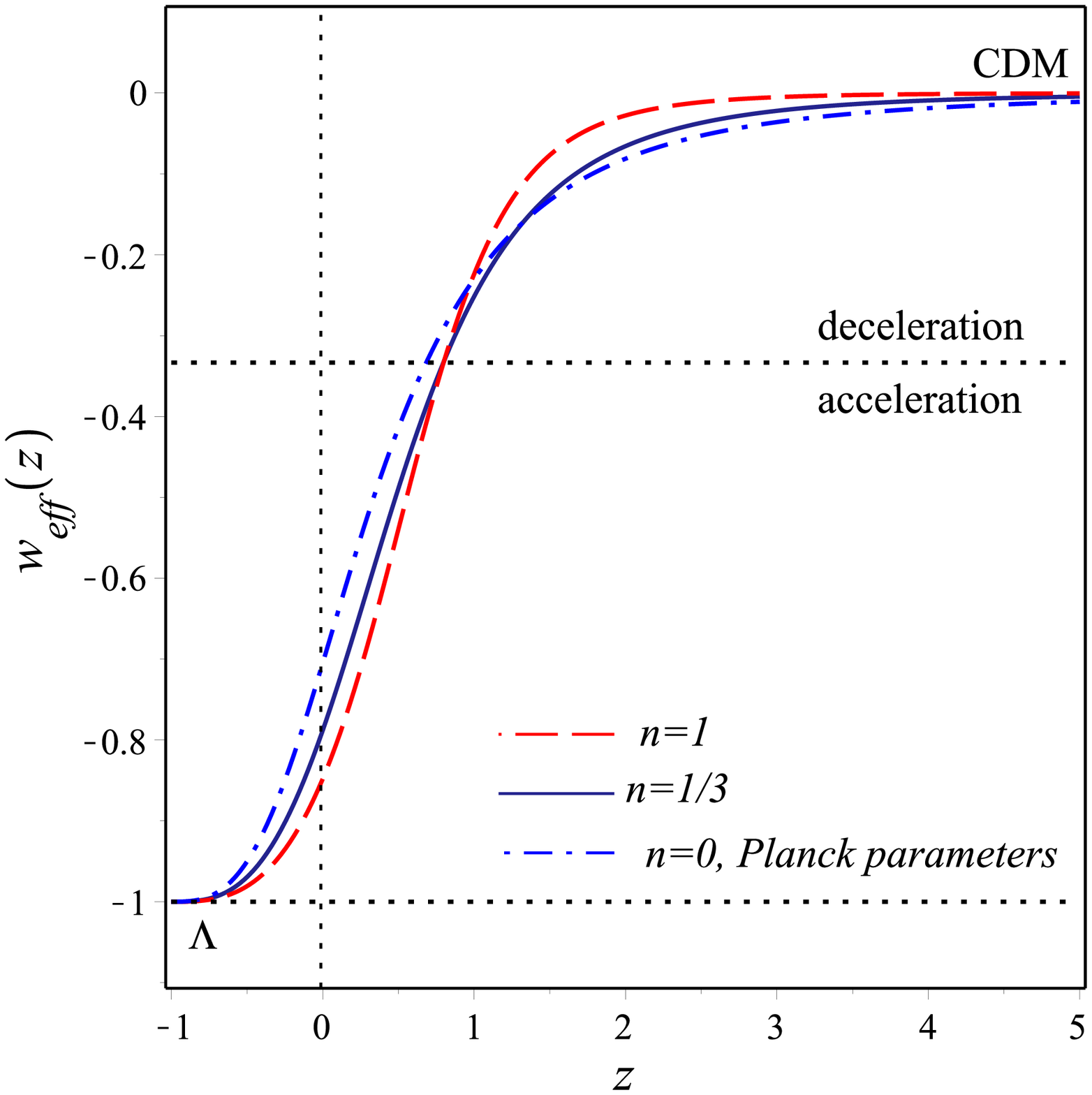}{0.3\textwidth}{}
          \fig{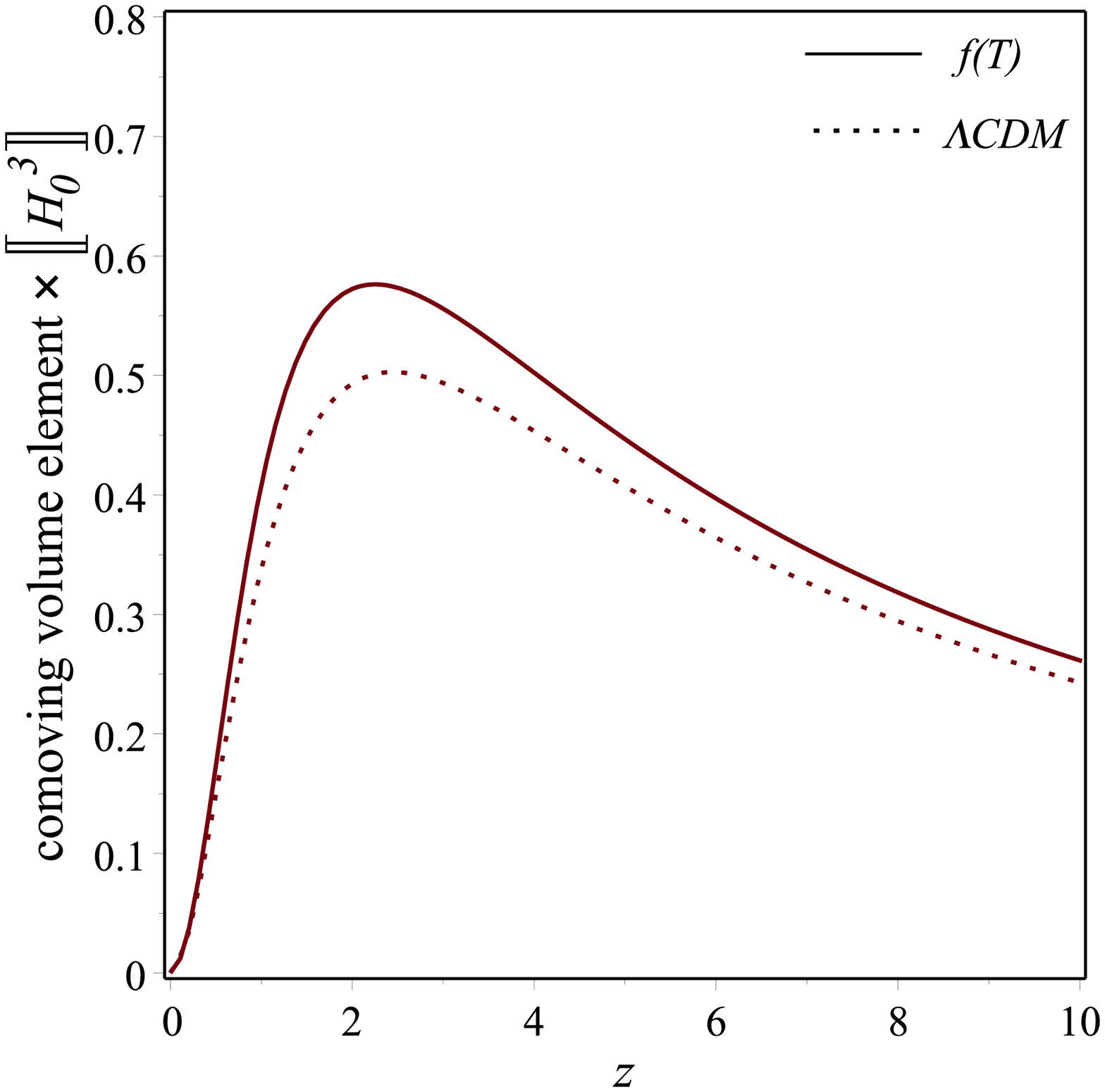}{0.3\textwidth}{}}
\caption{\textit{The cosmological parameters of the torsional IR correction. Left panel: The evolution of the matter and  the torsional density parameters, $\Omega_m(z)$ and $\Omega_T(z)$, from (\ref{matter-density-parameter}) and
(\ref{torsion-density-parameter}), respectively. Middle panel: The evolution of $w_{eff}$, namely (\ref{EoS_eff1}), shows that the universe effectively matches the CDM with $w_{eff}\to 0$ at past and evolves towards de Sitter $w_{eff}\to -1$ as $z\to -1$ with a transition redshift $z_{tr}=0.71$ as $w_{eff}=-1/3$. Right panel: The evolution of the volume element (\ref{volume_elemenet}) to a factor of $H_0^3$.} \label{Fig:Cosmological_parameters}}
\end{figure*}

\subsection{Cosmological parameters}\label{Sec:A.1}
For $n=1$ case, the torsion gravity model (\ref{fTn}) reads
\begin{equation}\label{fT}
    f(T)=T+\alpha \frac{~T_{0}^{2}}{T}.
\end{equation}
The modified Friedmanns' equations, (\ref{FR1T}) and (\ref{FR2T}), become
\begin{eqnarray}
  \rho &=& \frac{3}{\kappa^2}\left[H^2-3\alpha H_0^2\left(\frac{H_0}{H}\right)^2\right], \label{FR1H}\\
  p &=& -\frac{2}{\kappa^2}\dot{H}\left[1+3\alpha \left(\frac{H_0}{H}\right)^4\right]-\rho. \label{FR2H}
\end{eqnarray}

By constraining the above to the linear EoS choice $p=w \rho$, the solution is given as
\begin{eqnarray}\label{time}
\nonumber t&=&t_0+\frac{2}{3(1+w)H}\\
&+&\frac{3^{\frac{3}{4}}}{9}\frac{\left[\ln\left(\frac{H+(3\alpha)^{\frac{1}{4}}H_0}{H-(3\alpha)^{\frac{1}{4}}H_0}\right)
-2\arctan\left(\frac{H}{(3\alpha)^{\frac{1}{4}}H_0}\right)\right]}{(1+w)\alpha^{\frac{1}{4}} H_0},\label{time}
\end{eqnarray}
where $t_{0}$ is an integration constant. Although, the above solution is exact, it is hard to extract information about the system from (\ref{time}). For example, its not clear how the system could behave at $t\to \infty$, or how sensitive it is to the choice of initial conditions. On the contrary, as we have shown, the graphical analysis of its phase portrait represents an adequate description of the qualitative features of the global dynamics. For $n=1$ model, the phase portrait (\ref{phase_portrait}) reads
\begin{equation}\label{phase_portrait1}
    \dot{H}= -\frac{3}{2}(1+w)H^2\left[\frac{(H/H_0)^4-3\alpha}{(H/H_0)^4+3\alpha}\right],
\end{equation}
which has been drawn in Fig. \ref{Fig:phase_portrait}. As clear from (\ref{FR1H}) and (\ref{FR2H}) that the torsional counterpart has density and pressure,
\begin{eqnarray}
  \rho_T &=& \frac{9\alpha H_0^4}{\kappa^2 H^2}, \label{Torsion-density_H1}\\
  p_T &=& -\frac{18 \alpha H_0^4 H^2}{\kappa^2(H^4+3\alpha H_0^4)}. \label{Torsion-pressure_H1}
\end{eqnarray}

It is useful to represent the Friedmann equation (\ref{FR1H}) in  dimensionless form:
\begin{equation}\label{dimensionless_Friedmann}
    \Omega_{m}+\Omega_{T}=1,
\end{equation}
where $\Omega_{m}=\rho/\rho_{eff}$ and $\Omega_{T}=\rho_{T}/\rho_{eff}$ are the matter and the torsion density parameters, respectively. Also, we note that the model parameter $\alpha$, namely Eq. (\ref{alphan}), is related to current matter density parameter,
\begin{equation}\label{alpha}
    \alpha=\frac{1}{3}(1-\Omega_{m,0})=\frac{1}{3}\Omega_{T,0}.
\end{equation}
Using the above equation and the useful relation
\begin{equation}\label{Hdz}
\dot{H}=-(1+z) H(z) \frac{dH}{dz},
\end{equation}
one can solve (\ref{phase_portrait}) for Hubble
\begin{equation}\label{Hubble1}
    H(z)=\frac{H_0}{\sqrt{2}}\sqrt{\Omega_{m,0}(1+z)^3 + \sqrt{\Omega_{m,0}^2(1+z)^6+4\Omega_{T,0}}}.
\end{equation}
One of the important results which can be directly extracted from (\ref{Hubble1}) is the age-redshift relation.
\begin{equation}\label{age-redshift1}
    t(z)=\frac{\sqrt{2}}{H_0}\int_{z}^{\infty}\frac{dz'/(1+z')}{\sqrt{\Omega_{m,0}(1+z')^3+\sqrt{\Omega_{m,0}^2(1+z')^6+4\Omega_{T,0}}}}.
\end{equation}
Next we evaluate the matter density parameter by substituting from (\ref{Hubble1}) into (\ref{FR1H}), which yields
\begin{equation}\label{matter-density-parameter}
    \Omega_{m}(z)=\frac{2\,\Omega_{m,0}(1+z)^3}{\Omega_{m,0}(1+z)^3+\sqrt{\Omega_{m,0}^2(1+z)^6+4\Omega_{T,0}}}.
\end{equation}
Thus, the torsional density parameter is
\begin{equation}\label{torsion-density-parameter}
    \Omega_{T}(z)=1-\frac{2\,\Omega_{m,0}(1+z)^3}{\Omega_{m,0}(1+z)^3+\sqrt{\Omega_{m,0}^2(1+z)^6+4\Omega_{T,0}}}.
\end{equation}
We plot the evolution of $\Omega_m(z)$ and $\Omega_T(z)$ in Fig.~\ref{Fig:Cosmological_parameters} (left panel). It shows that $\Omega_m\to 1$ at large $z$ while $\Omega_T \to 0$, which indicates the CDM domination. On the contrary, $\Omega_m$ drops to zero and $\Omega_T\to 1$ at $z\to -1$ ($t\to \infty$), where the evolution is dominated with the dark torsion with a pseudo-rip cosmology as a final fate. The pattern shown in Fig.~\ref{Fig:Cosmological_parameters} (left panel) is in agreement with basic requirements of the viable scenario.

Using Eqs. (\ref{Hdz})  and (\ref{Hubble1}), the deceleration parameter of the torsional IR model is given by
\begin{equation}\label{deceleration1}
    q(z)=-1+\frac{3\Omega_{m,0}(1+z)^3}{2\sqrt{\Omega_{m,0}^2(1+z)^6+4\Omega_{T,0}}}.
\end{equation}
Alternatively, using (\ref{deceleration}), we write the effective (total) EoS
\begin{equation}\label{EoS_eff1}
    w_{eff}(z)=-1+\frac{\Omega_{m,0}(1+z)^3}{\sqrt{\Omega_{m,0}^2(1+z)^6+4\Omega_{T,0}}},
\end{equation}
which is plotted as in Fig. \ref{Fig:Cosmological_parameters}(middle panel), it shows that $-1\geq w_{eff} \leq -1/3$ at $-1 \geq z \lesssim 0.7$ in agreement with observations. However, to express the torsional counterpart EoS in terms of redshift, $w_T(z)$, we substitute (\ref{Hubble1}) into (\ref{Torsion-density_H}) and (\ref{Torsion-pressure_H}), to write its density and pressure
\begin{eqnarray}
\nonumber \rho_{T}(z)&=&\frac{6\Omega_{T,0}H_0^2}{\kappa^2\left[\Omega_{m,0}(1+z)^3+\sqrt{\Omega_{m,0}^2(1+z)^6+4\Omega_{T,0}}~\right]},\label{Torsion-density1}\\
    p_{T}(z)&=&-\frac{6\Omega_{T,0}H_0^2}{\kappa^2 \sqrt{\Omega_{m,0}^2(1+z)^6+4\Omega_{T,0}}}.\label{Torsion-pressure1}
\end{eqnarray}
Hence, we obtain the torsional EoS
\begin{equation}\label{EoS_Torsion1}
    w_{T}(z)=-1-\frac{\Omega_{m,0}(1+z)^3}{\sqrt{\Omega_{m,0}^2(1+z)^6+4\Omega_{T,0}}}.
\end{equation}
At present, $z=0$, the above equation reduces to
\[w_{T,0}=-1-\frac{\Omega_{m,0}}{2-\Omega_{m,0}}.\]
For any value $\Omega_{m,0}>0$, the torsional EoS goes below $-1$. This clarifies the phantom-like nature of the torsional IR corrections.
Also, we note that the angular distance, namely (\ref{ang_dist}), allows to perform an important qualitative test, that is the evolution of the comoving volume element within solid angle $d\Omega$ and redshift $dz$,
\begin{equation}\label{volume_elemenet}
    dV=\frac{(1+z)^2 D_A^2}{H(z)}d\Omega \, dz.
\end{equation}
This quantity provides a useful test for computing the source counts \cite{Newman:1999cg}. Using (\ref{Hubble1}) and (\ref{ang_dist}), the evolution of the volume element (up to a factor of Hubble volume $H_0^3$) is plotted in Fig.~\ref{Fig:Cosmological_parameters} (right panel). The plot shows that the comoving volume element reaches a maximum value at $z \gtrsim 2$ very similar to the $\Lambda$CDM pattern.
\subsection{Physical Viability}\label{Sec:A.2}
In addition, we perform a basic test on the perturbation level of the theory which should be carried out for any modified gravity theory, that is the propagation of the sound speed of the scalar fluctuations. As a matter of fact a considerable array of modified gravity theories can describe the late transition of the cosmic acceleration fulfilling the basic requirements on the background level. However, any such theory remains  at risk until its description on the perturbation level too fulfills some physical conditions. A necessary condition is for the sound speed of scalar fluctuations to be constrained between  $0\leq c_s^2 \leq 1$. This is required in order to have a stable and causal theory.
\begin{figure}
\begin{center}
\includegraphics[scale=0.3]{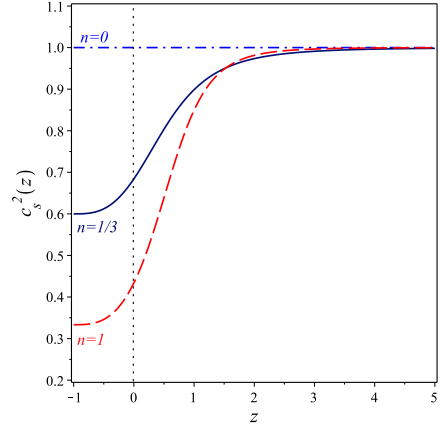}
\caption{\textit{The evolution of the square of the sound speed of the scalar fluctuations (\ref{sound_speed_z}). The plots show that $c_s^2(z) \to 1$ as $z \to \infty$, for all $n$ values, just as in $\Lambda$CDM model, so we do not expect any deviation on the perturbation level of the theory from the standard cosmology. However, for $n=1$ ($n=1/3$) models, they evolve towards $c_s^2(z) \to \frac{1}{3}$ ($0.6$) as $z \to -1$ at far future, respectively. This may have some impacts on the modern galaxy formation. In general, the theory is in agreement with the stability and causality conditions as $0 \leq c_s^2 \leq 1$. We use $\Omega_{m,0}=0.262$ and $H_0=73.5$ km/s/Mpc.}}
\label{Fig:sound_speed}
\end{center}
\end{figure}

To calculate the sound speed we  take the longitudinal gauge with two scalars metric fluctuation, that is
\begin{equation}\label{pert-metric}
    ds^2=(1+2\Phi)dt^2-a^2(1-2\Psi)dx^2.
\end{equation}
This leads to a fluctuation in the teleparallel torsion scalar \cite{Cai:2011tc}
\[\delta T = 12 H (\dot{\Phi}+H\Psi).\]
Just as in  GR theory, the weak field limit about Minkowski space clarifies that the scalar metric fluctuation $\Phi$ plays the role of the gravitational potential. We follow the perturbation equations \cite{Cai:2011tc} up to the linear order, assuming the matter sector is a canonical scalar field $\phi$ with a lagrangian
\begin{equation}\label{lag-scalar-field}
    \mathcal{L}_{m}\to \mathcal{L}_{\phi}=\frac{1}{2}\partial_{\mu}\phi \, \partial^{\mu}\phi-V(\phi).\vspace{0.1 cm}
\end{equation}

For the choice of the vierbein (\ref{tetrad}), it has been shown that (see \cite{Chen:2010va,Cai:2011tc}), in the $f(T)$ gravity, we have only a single degree of freedom minimally coupled to a canonical scalar field $\phi$, since the scalar field fluctuation $\delta \phi$ can fully determine the gravitational potential $\Phi$ in the absence of anisotropic stress, i.e $\Phi=\Psi$. Using the relation (\ref{TorHubble}), we find that the square of the sound speed\footnote{Usually the square of the sound speed of the scalar fluctuations is given in the form $c_{s}^{2}=\frac{f_T}{f_T+2T f_{TT}}$ (see \cite{Chen:2010va,Cai:2011tc}). We reexpress it in terms of $H$ as given in Eq. (\ref{sound_speed_H}), which is more appropriate for our analysis.} for the general form of the IR $f(T)$ theory (\ref{fTn}),
\begin{equation}\label{sound_speed_H}
    c_{s}^{2}=\frac{f_{H}}{H f_{HH}}=1-\frac{2n(1+n)(1-\Omega_{m,0})}{(1+2n)[\tilde{H}^{2(n+1)}+n(1-\Omega_{m,0})]}.
\end{equation}
As clear, for $n=0$, the model reduces to $\Lambda$CDM where the speed of sound is fixed to the value $c_s=1$. For $n=1$ case, we substitute from (\ref{Hubble1}) into (\ref{sound_speed_H}), we write the square of the sound speed in terms of the redshift,
\begin{equation}\label{sound_speed_z}
    c_{s}^{2}(z)=1-\frac{8\Omega_{T,0}/\sqrt{\Omega_{m,0}^2(1+z)^6+4\Omega_{T,0}}}
    {3\,\left(\Omega_{m,0}(1+z)^3+\sqrt{\Omega_{m,0}^2 (1+z)^6 +4\Omega_{T,0}}\right)}.
\end{equation}

We thus can verify that the square of the sound speed of the primordial scalar fluctuation $c_s^2\to 1$ at past as $z\to \infty$, while its current value $c_{s}^{2}(z=0)\sim 0.43$. However, at the far future $c_s^2\to \frac{1}{3}$ as $z\to -1$. The detailed evolution is given in Fig.~\ref{Fig:sound_speed}, which shows that the square of the sound speed is $\frac{1}{3}\leq c_s^2 \leq 1$. Also, we include the evolution of $c_s^2(z)$ in the ($n=1/3$) case for completeness. This result confirms that the torsional IR correction theory is free from ghost/gradiant instabilities and acausality problems at all times.

\end{document}